\documentclass[final,5p,times,twocolumn]{elsarticle}

\usepackage[american]{babel}
\usepackage{amssymb}
\usepackage{amsmath}
\usepackage{graphicx}
\usepackage{subcaption}
\usepackage{siunitx}
\usepackage{physics}
\usepackage{pdfrender}
\usepackage{comment}
\usepackage[thinc]{esdiff}
\usepackage{lipsum}
\usepackage{amsthm}
\usepackage[capitalise]{cleveref}
\usepackage{multirow, makecell}
\usepackage{nicefrac}
\usepackage{pifont}
\usepackage{scalerel}
\usepackage{mathtools}
\usepackage{booktabs}

\usepackage{lineno}

\makeatletter
\renewcommand*\env@matrix[1][\arraystretch]{%
  \edef\arraystretch{#1}%
  \hskip -\arraycolsep
  \let\@ifnextchar\new@ifnextchar
  \array{*\c@MaxMatrixCols c}}
\makeatother

\makeatletter
\newcommand*\rel@kern[1]{\kern#1\dimexpr\macc@kerna}
\newcommand*\widebar[1]{%
  \begingroup
  \def\mathaccent##1##2{%
    \rel@kern{0.8}%
    \overline{\rel@kern{-0.8}\macc@nucleus\rel@kern{0.2}}%
    \rel@kern{-0.2}%
  }%
  \macc@depth\@ne
  \let\math@bgroup\@empty \let\math@egroup\macc@set@skewchar
  \mathsurround\z@ \frozen@everymath{\mathgroup\macc@group\relax}%
  \macc@set@skewchar\relax
  \let\mathaccentV\macc@nested@a
  \macc@nested@a\relax111{#1}%
  \endgroup
}

\newcommand{\defaultstretch}{1.25}
\renewcommand{\arraystretch}{\defaultstretch}

\newcommand*{\boldgreek}[1]{%
  \textpdfrender{%
    TextRenderingMode=FillStroke,%
    LineWidth=.45pt,%
  }{#1}%
}

\newcommand{\matrixSpacingDefault}{1.4}
\newcommand{\matrixSpacingBig}{2.1}
\newcommand{\pdiffScaleFactor}{1.35}
\newcommand{\lastRowMargin}{3pt}
\newcommand{\mquad}{\;\,}

\newcommand{\pdiff}[2]{\ensuremath{\mathchoice
{\frac{\partial{#1}}{\partial{#2}}}%
{\frac{\partial{#1}}{\partial{#2}}}%
{\nicefrac{\partial{#1}}{\partial{#2}}}%
{\nicefrac{\partial{#1}}{\partial{#2}}}}}
\newcommand{\pdiffdiff}[2]{\ensuremath{\frac{\partial^2{#1}}{\partial{ #2}^2}}}

\newcommand{\initial}{\mathtt{I}}
\newcommand{\final}{\mathtt{F}}

\newcommand{\covariance}{\mathbf{C}}
\newcommand{\covarianceInitial}{\covariance_\initial}
\newcommand{\covarianceFinal}{\covariance_\final}
\newcommand{\covarianceInitialReference}{\covarianceInitial}

\newcommand{\jacobianGeneral}{\mathbf{J}}
\newcommand{\jacobianFull}{\jacobianGeneral}
\newcommand{\jacobianLocalToGlobal}{\jacobianGeneral_{L \rightarrow G}}
\newcommand{\jacobianGlobalToLocal}{\jacobianGeneral_{G \rightarrow L}}
\newcommand{\jacobianGlobalToGlobal}{\jacobianGeneral_{T}}

\newcommand{\azimu}{\phi}
\newcommand{\azimuInitial}{\azimu_\initial}
\newcommand{\azimuFinal}{\azimu_\final}
\newcommand{\polar}{\theta}
\newcommand{\polarInitial}{\polar_\initial}
\newcommand{\polarFinal}{\polar_\final}

\newcommand{\pos}{\mathbf{r}}
\newcommand{\posInitial}{\pos_{\initial}}
\newcommand{\posFinal}{\pos_{\final}}
\newcommand{\dirC}{n}
\newcommand{\dir}{\mathbf{\dirC}}
\newcommand{\dirX}{\dirC_x}
\newcommand{\dirY}{\dirC_y}
\newcommand{\dirZ}{\dirC_z}
\newcommand{\dirInitial}{\dir_{\initial}}
\newcommand{\dirFinal}{\dir_{\final}}
\newcommand{\qop}{\lambda}
\newcommand{\qopInitial}{\qop_{\initial}}
\newcommand{\qopFinal}{\qop_{\final}}
\newcommand{\bfield}{\mathbf{b}}
\newcommand{\bfieldFinal}{\bfield_{\final}}
\newcommand{\pathlength}{s}
\newcommand{\charge}{q}
\newcommand{\momentum}{p}

\newcommand{\energy}{E}
\newcommand{\energyFinal}{\energy_{\final}}

\newcommand{\local}{\boldgreek{\rho}}
\newcommand{\localInitial}{\local_\initial}
\newcommand{\localFinal}{\local_\final}
\newcommand{\sphAngle}{\boldgreek{\omega}}
\newcommand{\sphAngleInitial}{\sphAngle_\initial}
\newcommand{\sphAngleFinal}{\sphAngle_\final}
\newcommand{\frameCenter}{\mathbf{c}}
\newcommand{\planeNormal}{\mathbf{w}}
\newcommand{\planeNormalFinal}{\planeNormal_\final}
\newcommand{\ubasis}{\mathbf{u}}
\newcommand{\vbasis}{\mathbf{v}}
\newcommand{\ubasisInitial}{\ubasis_\initial}
\newcommand{\vbasisInitial}{\vbasis_\initial}
\newcommand{\ubasisFinal}{\ubasis_\final}
\newcommand{\vbasisFinal}{\vbasis_\final}
\newcommand{\ulocal}{l_0}
\newcommand{\ulocalInitial}{l_{0\initial}}
\newcommand{\ulocalFinal}{l_{0\final}}
\newcommand{\vlocal}{l_1}
\newcommand{\vlocalInitial}{l_{1\initial}}
\newcommand{\vlocalFinal}{l_{1\final}}
\newcommand{\vtangle}{\zeta}

\newcommand{\transpose}[1]{\ensuremath{#1^{\mathsf{T}}}}
\newcommand{\inverse}[1]{\ensuremath{#1^{-1}}}

\newcommand{\diffquote}[1]{\ensuremath{#1'}}
\newcommand{\doubquote}[1]{\ensuremath{#1''}}

\newcommand{\stoppingPower}{-\frac{\partial \energy}{\partial \pathlength}}

\newcommand{\drdphiInitial}{
\pdiff{\posInitial}{\azimuInitial} = \frac{\ulocalInitial}{|\vbasisInitial \times \dirInitial|} \left[ - \ubasisInitial \left(  \vbasisInitial \times \pdiff{\dirInitial}{\azimuInitial} \cdot \ubasisInitial \right) + \vbasisInitial \times \pdiff{\dirInitial}{\azimuInitial}  \right]
}

\newcommand{\drdphi}{
\pdiff{\pos}{\azimu} = \frac{\ulocal}{|\vbasis \times \dir|} \left[ - \ubasis \left(  \vbasis \times \pdiff{\dir}{\azimu} \cdot \ubasis \right) + \vbasis \times \pdiff{\dir}{\azimu}  \right]
}

\newcommand{\atomicNumber}{Z}
\newcommand{\massNumber}{A}
\newcommand{\mass}{m}
\newcommand{\Kcoeff}{K}
\newcommand{\meanExcitE}{I}
\newcommand{\maxTransE}{W_{\textrm{max}}}
\newcommand{\radiationLength}{X_0}

\newcommand{\zeroMatrix}[2]{\mathbb{O}_{#1#2}}
\newcommand{\identityMatrix}[1]{\mathbf{I}_{#1}}
\newcommand{\perigeeAdditionalTerm}{\kappa}
\newcommand{\indexI}{i}
\newcommand{\indexJ}{j}
\newcommand{\indexK}{k}
\newcommand{\indexL}{l}
\newcommand{\indexRK}{n}
\newcommand{\indexNewton}{n}
\newcommand{\scalarFunc}{f}
\newcommand{\implicitF}{F}
\newcommand{\symmetricQuotientN}{N}

\newcommand{\boundImplicit}{\implicitF}
\newcommand{\perigeeImplicit}{\implicitF}
\newcommand{\InitialToFinalMapping}{G}
\newcommand{\stdbasis}{\mathbf{e}}
\newcommand{\choleskydecompose}{\mathbf{L}}
\newcommand{\normalrandom}{\boldgreek{\eta}}
\newcommand{\residual}{\boldgreek{\delta}}
\newcommand{\globalPars}{\mathbf{g}}
\newcommand{\globalParsInitial}{\globalPars_\initial}
\newcommand{\globalParsFinal}{\globalPars_\final}
\newcommand{\localPars}{\boldgreek{\xi}}
\newcommand{\localParsInitial}{\localPars_\initial}
\newcommand{\localParsInitialSmeared}{\tilde{\localParsInitial}}
\newcommand{\localParsFinal}{\localPars_\final}
\newcommand{\localParsFinalPlus}{\localPars_{\final+}}
\newcommand{\localParsFinalMinus}{\localPars_{\final-}}
\newcommand{\localParsInitialReference}{\localParsInitial}
\newcommand{\localParsFinalReference}{\localParsFinal}
\newcommand{\localParsFinalSmeared}{\tilde{\localParsFinal}}
\newcommand{\dpl}{\epsilon}
\newcommand{\unitVector}{\mathbf{e}}
\newcommand{\unitZ}{\unitVector_{z}}
\newcommand{\correlation}{\Gamma_c}

\newcommand{\stepsize}{h}
\newcommand{\ddirds}{\mathbf{k}}
\newcommand{\dqopds}{\diffquote{\qop}}
\newcommand{\globalError}{\tau}
\newcommand{\localError}{\abs{\Delta\newstate{\pos}}}
\newcommand{\newstate}[1]{\ensuremath{\bar{#1}}}
\newcommand{\jacobianStep}{\mathbf{S}}
\newcommand{\equationSpacingA}{0.2em}
\newcommand{\equationSpacingB}{0.6em}
\newcommand{\nTracks}{{10^5}}
\newcommand{\nFourSigma}{{6.33}}
\newcommand{\logToleranceMin}{-18}
\newcommand{\logToleranceMax}{6}
\newcommand{\logToleranceLimit}{-20}
\newcommand{\logToleranceShiftedTracks}{-16}
\newcommand{\logToleranceCovTransport}{-8}
\newcommand{\toleranceMin}{10^{\logToleranceMin}}
\newcommand{\toleranceMax}{10^{\logToleranceMax}}
\newcommand{\toleranceLimit}{10^{\logToleranceLimit}}
\newcommand{\toleranceShitedTracks}{10^{\logToleranceShiftedTracks}}
\newcommand{\toleranceCovTransport}{10^{\logToleranceCovTransport}}
\newcommand{\chargeNumber}{z_c}
\newcommand{\incidentMass}{M}

\newcommand{\logbetagamma}{\mu}
\newcommand{\density}{\rho_m}
\newcommand{\helicalDir}{\dir_w}

\newcommand{\absRelResidual}{\Omega_R}
\newcommand{\pullSymbol}{\mathrm{PL}}
\newcommand{\pull}[1]{\pullSymbol(#1)}
\newcommand{\numDiffMatrix}{\mathbf{D}}
\newcommand{\curvilinearVector}{\mathbf{e}_c}
\newcommand{\columnwiseCrossProd}{\otimes}
\newcommand{\RKstepsTotal}{N}
\newcommand{\leftBraIdx}{[}
\newcommand{\rightBraIdx}{]}
\newcommand{\pvalue}{p}

\newcommand{\globalParsZero}{\globalPars_0}
\newcommand{\posZero}{\pos_0}
\newcommand{\dirZero}{\dir_0}
\newcommand{\globalParsHelix}{\globalPars_\pathlength}
\newcommand{\posHelix}{\pos_\pathlength}
\newcommand{\dirHelix}{\dir_\pathlength}
\newcommand{\qopHelix}{\qop}
\newcommand{\bfieldNormalized}{\hat{\bfield}}
\newcommand{\QtermHelix}{Q}
\newcommand{\riddersMatrix}{\mathbf{U}}
\newcommand{\riddersMatrixIJ}{\mathbf{U}\textsuperscript{$\indexI\indexJ$}}
\newcommand{\riddersRate}{R}
\newcommand{\riddersIteration}{N}
\newcommand{\riddersCompleteIndex}{n}

\newcommand{\frameFigureSize}{0.82}
\newcommand{\rkStepFigureSize}{0.82}
\newcommand{\rkToleranceFigureSize}{0.41}
\newcommand{\histogramFigureSize}{0.82}
\newcommand{\jaocobiCompareFigureSize}{0.82}
\newcommand{\geometryFigureSize}{0.41}
\newcommand{\pullpvalFigureSize}{0.41}

\DeclareSIUnit\clight{\text{\ensuremath{c}}}

\journal{Nuclear Instruments and Methods in Physics Research Section A}

\begin{document}

\begin{frontmatter}




\title{The derivation of Jacobian matrices for the propagation of track parameter uncertainties in the presence of magnetic fields and detector material}


\author[UCB,LBNL]{Beomki Yeo\corref{cor1}}
\author[UCB,LBNL]{Heather Gray}
\author[CERN]{Andreas Salzburger}
\author[CERN,UOA]{Stephen Nicholas Swatman}
\cortext[cor1]{beomki.yeo@berkeley.edu}
\affiliation[UCB]{
            organization={Department of Physics, University of California}, 
            city={Berkeley},
            postcode={CA 94720}, 
            country={USA}}
\affiliation[LBNL]{
            organization={Physics Division, Lawrence Berkeley National Laboratory}, city={Berkeley},
            postcode={CA 94720},
            country={USA}}
\affiliation[CERN]{
            organization={European Organization for Nuclear Research}, city={Meyrin},
            postcode={1211},
            country={Switzerland}}            
\affiliation[UOA]{
            organization={University of Amsterdam}, city={Amsterdam},
            postcode={1012 WX},
            country={The Netherlands}}
\begin{abstract}

In high-energy physics experiments, the trajectories of charged particles are reconstructed using track reconstruction algorithms. Such algorithms need to both identify the set of measurements from a single charged particle and to fit the parameters by propagating tracks along the measurements. The propagation of the track parameter uncertainties is an important component in the track fitting to get the optimal precision in the fitted parameters. The error propagation is performed at intersections between the track and local coordinate frames defined on detector components by calculating a Jacobian matrix corresponding to the local-to-local frame transport. This paper derives the Jacobian matrix in a general manner to harmonize with numerical integration methods developed for inhomogeneous magnetic fields and materials. The Jacobian and transported covariance matrices are validated by simulating the propagation of charged particles between two frames and comparing with the results of numerical methods.


\end{abstract}




\end{frontmatter}




\section{Introduction}
\label{sec:introduction}

In high-energy physics experiments, charged particles traverse detectors permeated by magnetic fields and undergo electromagnetic interactions with the detector material. The reconstruction of the trajectories of the particles---known also as tracks---is performed by identifying a set of measurements corresponding to a single charged particle and fitting initial track parameters along those measurements. The fitting process requires spatial information about the measurements, as well as an estimate of the initial track parameters, and even better results can be achieved by considering measurement noises and track parameter uncertainties derived from physics models. The uncertainties of track parameters at each point associated with a measurement can be obtained using error propagation. This requires the derivative of a model function to be taken with respect to function variables---i.e., a Jacobian matrix---if the physics models, like those in track propagation, are non-linear.

Kalman filtering~\cite{KALMAN1960_Kalman, FRUHWIRTH1987_Kalman, FRUHWIRTH2021_PatternRecognition} and least-squares regression~\cite{FRUHWIRTH2021_PatternRecognition} are the most popular track fitting methods used to perform error propagation with Jacobian matrices. Kalman filtering  iteratively proceeds along the measurements to propagate the track parameters and transport their covariance matrix to the surfaces where the measurements are defined. The errors of propagated track parameters are minimized by updating them based on the covariance matrix, measurement and its noise. The aforementioned covariance matrix transport is followed by the evaluation of a Jacobian matrix which is a derivative of the track parameters at the local coordinate frame of the current surface with respect to those at the previous surface. Least-squares regression defines an estimator for initial track parameters which takes into account all measurements collectively and utilizes the Jacobian matrices to update the initial track parameters iteratively until the corresponding chi-square is converged.

Previous work~\cite{WITTEK1980_Helix, WITTEK1981_Helix, STRANDLIE2006_Helix} derives Jacobian matrices analytically for helices in homogeneous magnetic fields. However, this method cannot be straightforwardly extended to non-helical trajectories in inhomogeneous magnetic fields, which do not have analytical solutions. Previous work~\cite{BUGGE1981_RK, LUND2009_CovarianceTransport} also provides numerical methods for the transport of covariance matrices in inhomogeneous magnetic fields and materials, but these methods do not address constraints arising from the intersection condition between the track and the local coordinate frame during the calculation of the Jacobian matrix. 

This paper extends the scope of previous studies~\cite{WITTEK1980_Helix, WITTEK1981_Helix, STRANDLIE2006_Helix, BUGGE1981_RK, LUND2009_CovarianceTransport} to derive a Jacobian matrix for covariance matrix transport between local coordinate frames in inhomogeneous magnetic fields and materials, which can be implemented with arbitrary numerical integration models such as the Runge-Kutta-Nystr\"{o}m (RKN) method~\cite{BUGGE1981_RK, LUND2009_CovarianceTransport, LUND2009_RKNPropagation, NYSTROM1925_RKN}. Although Jacobian matrices can manifest differently depending on the selection of the local coordinate frame, we study two representative frame types defined in the Cartesian coordinate system: one is the bound frame which is a local coordinate frame defined on a plane (typically a detector element) and the other is the perigee frame~\cite{BILLOIR1992_Perigee} which is not bound to a plane\footnote{The perigee frame is important for vertex reconstruction and track parametrization with wire measurements in detectors containing drift chambers and straw tubes.}. 

The paper is organized as follows: \cref{sec:notation} introduces mathematical notation. \cref{sec:derivation} derives Jacobian matrices for  bound and perigee frames. \cref{sec:validation} validates the Jacobian matrices against numerical differentiation and examines if the statistics of transported track parameters follow the transported covariance matrix. \cref{sec:discussion} discusses possible improvements for the experimental implementation. Finally, \cref{sec:summary} provides a summary of our work. 

\section{Notation}\label{sec:notation}
Scalars are represented by variables with an italic font regardless of the letter case, e.g., $\alpha$ and $T$. Vectors and matrices, both in a bold typeface, are denoted as lowercase and uppercase letters, respectively, except the $m \times n$ zero matrix which is denoted as $\zeroMatrix{m}{n}$. We use subscripts in square brackets to denote individual elements of a vector or a matrix. For example, $\mathbf{y}_{\leftBraIdx\indexI\rightBraIdx}$ is the $\indexI$-th element of vector $\mathbf{y}$, and $\mathbf{A}_{\leftBraIdx\indexI,\indexJ\rightBraIdx}$ is the element at the $\indexI$-th row and the $\indexJ$-th column of matrix $\mathbf{A}$.

As a convention for a vector-by-vector derivative such as a Jacobian matrix, we follow a so-called numerator layout: $\pdiff{\mathbf{y}}{\mathbf{x}}$ is represented as a matrix where the elements of $\mathbf{x}$ and $\mathbf{y}$ are expanded in columns and rows, respectively. 

The inner product of two vectors is denoted by $\mathbf{x} \cdot \mathbf{y}$ and the cross product by $\mathbf{x} \times \mathbf{y}$. $\transpose{\mathbf{A}}$ is the transpose matrix of $\mathbf{A}$, and $\inverse{\mathbf{A}}$ is the inverse matrix of $\mathbf{A}$. $\identityMatrix{n}$ is defined as the $n\times n$ identity matrix.

\section{Mathematical derivation} \label{sec:derivation}

The movement of charged particles in a magnetic field is governed by the Lorentz force, which can be described by the following second-order ordinary differential equation:

\begin{equation}\label{eq:EquationOfMotion}
    \pdiffdiff{\pos}{\pathlength} = \frac{\charge}{\momentum} \left( \pdiff{\pos}{\pathlength} \times \bfield \right) = \qop \left( \dir  \times \bfield \right),
\end{equation}
where $\pos$ is the three-dimensional global Cartesian position $\transpose{(x\mquad y\mquad z)}$, $\pathlength$ is the path length along the trajectory, and $\dir$ is the three-dimensional unit tangential direction $\transpose{(\dirX\mquad \dirY\mquad \dirZ)}$, equivalent to $\pdiff{\pos}{\pathlength}$. $\bfield$ is the magnetic field, which depends on $\pos$ for inhomogeneous magnetic fields. $\charge$ and $\momentum$ are the charge and momentum of the particle, and $\qop$ is defined as $\charge/\momentum$, which depends on $\pathlength$ inside the detector material due to the energy loss. 

Tracks in space can be parametrized with seven global track parameters $\globalPars$ consisting of $\pos$,  $\dir$, and $\qop$: 
\begin{equation}
    \transpose{\globalPars} = 
    \begin{pmatrix}
        \transpose{\pos} &  \transpose{\dir}  & \qop
    \end{pmatrix} =
    \begin{pmatrix}
        x & y & z & \dirX & \dirY & \dirZ & \qop
    \end{pmatrix}, \\
\end{equation}
where the number of degrees of freedom of $\dir$ is two due to the unit vector condition. Therefore, the total number of degrees of freedom of the global track parameters is six.

Tracks on local coordinate frames can be parametrized with five local track parameters $\localPars$ consisting of $\local$, $\sphAngle$, and $\qop$:
\begin{equation}
    \transpose{\localPars} =
    \begin{pmatrix}
         \transpose{\local} &  \transpose{\sphAngle} & \qop
    \end{pmatrix} =
    \begin{pmatrix}
         \ulocal & \vlocal & \azimu & \polar & \qop
    \end{pmatrix},
\end{equation}
where $\local$ is the two-dimensional position $\transpose{(\ulocal \mquad \vlocal)}$ in the local Cartesian coordinate system, and $\sphAngle$ are the azimuthal and polar angles $\transpose{(\azimu\mquad  \polar)}$ in the global spherical coordinate system. As a convention, $\azimu$ is defined in [$-\pi$, $\pi$], and $\polar$ in [0, $\pi$]. The total number of degrees of freedom of the local track parameters is five as one spatial dimension is reduced by the constraint of a frame intersection. 

At a frame intersection where the number of degrees of $\globalPars$ is reduced to five, the components of $\globalPars$ can be represented with those of $\localPars$ and vice versa. The position, $\pos$ at the frame intersection can be represented as the following equation:
\begin{equation}\label{eq:r_in_uv}
\pos = \ulocal \ubasis + \vlocal \vbasis + \frameCenter,
\end{equation}
where $\ubasis$ and $\vbasis$ are the orthonormal basis of the local Cartesian coordinate system. $\frameCenter$ is the vector from the origin of the global coordinate system to the origin of the local coordinate frame: the origins
are where $\pos$ and $\local$ are zero vectors, respectively, and $\frameCenter$ can be chosen arbitrarily as long as the condition for the frame intersection is met, which depends on the frame type. It is also possible to represent the local position at the frame intersection as a function of the orthonormal basis, which is directly derived from \cref{eq:r_in_uv}:
\begin{equation}
\label{eq:mu_in_uv}
\local = 
\begin{pmatrix}[\matrixSpacingDefault]
  \ulocal \\
  \vlocal
\end{pmatrix} =
\begin{pmatrix}[\matrixSpacingDefault]
  (\pos - \frameCenter) \cdot \ubasis \\
  (\pos - \frameCenter) \cdot \vbasis 
\end{pmatrix}.
\end{equation}
$\dir$ and $\sphAngle$ can also represent each other as follows:
\begin{align}\label{eq:n_w_relations}
    \transpose{\dir} & = 
    \begin{pmatrix}
        \cos{\azimu}\sin{\polar} & \sin{\azimu}\sin{\polar} &
        \cos{\polar}
    \end{pmatrix}, \nonumber \\
    \transpose{\sphAngle} & =
    \begin{pmatrix}
        \arctan{(\dirY/\dirX)} & 
        \arctan{\left(\sqrt{\dirX^2+\dirY^2}/\dirZ\right)}
    \end{pmatrix}. 
\end{align}

Due to the non-linearity of \cref{eq:EquationOfMotion}, the covariance matrix of the track parameters can be updated with a Jacobian matrix by assuming that the system can be linearized with a first-order Taylor expansion. The error propagation for the local-to-local frame transport, which is the main purpose of this paper, is summarized as the following covariance matrix transport: 
\begin{equation}\label{eq:covariance_transport}
\covarianceFinal = \jacobianFull \covarianceInitial \transpose{\jacobianFull},
\end{equation}
where $\covarianceInitial$ and $\covarianceFinal$ are the covariance matrices of the local track parameters at the initial frame ($\localParsInitial$) and the final frame ($\localParsFinal$), respectively. $\jacobianFull$ is the Jacobian matrix that propagates the covariance matrices which is given as a $5\times5$ matrix: 
\begin{equation}\label{eq:full_jacobi_general}
 \jacobianFull = 
        \pdiff{\localParsFinal}{\localParsInitial} = 
        \begin{pmatrix}[\matrixSpacingBig]
        \scalebox{\pdiffScaleFactor}{\pdiff{\ulocalFinal}{\ulocalInitial}} &
        \scalebox{\pdiffScaleFactor}{\pdiff{\ulocalFinal}{\vlocalInitial}} &
        \scalebox{\pdiffScaleFactor}{\pdiff{\ulocalFinal}{\azimuInitial}} &
        \scalebox{\pdiffScaleFactor}{\pdiff{\ulocalFinal}{\polarInitial}} &
        \scalebox{\pdiffScaleFactor}{\pdiff{\ulocalFinal}{\qopInitial}} \\
        \scalebox{\pdiffScaleFactor}{\pdiff{\vlocalFinal}{\ulocalInitial}} &
        \scalebox{\pdiffScaleFactor}{\pdiff{\vlocalFinal}{\vlocalInitial}} &
        \scalebox{\pdiffScaleFactor}{\pdiff{\vlocalFinal}{\azimuInitial}} &
        \scalebox{\pdiffScaleFactor}{\pdiff{\vlocalFinal}{\polarInitial}} &
        \scalebox{\pdiffScaleFactor}{\pdiff{\vlocalFinal}{\qopInitial}} \\
        \scalebox{\pdiffScaleFactor}{\pdiff{\azimuFinal}{\ulocalInitial}} &
        \scalebox{\pdiffScaleFactor}{\pdiff{\azimuFinal}{\vlocalInitial}} &
        \scalebox{\pdiffScaleFactor}{\pdiff{\azimuFinal}{\azimuInitial}} &
        \scalebox{\pdiffScaleFactor}{\pdiff{\azimuFinal}{\polarInitial}} &
        \scalebox{\pdiffScaleFactor}{\pdiff{\azimuFinal}{\qopInitial}} \\
        \scalebox{\pdiffScaleFactor}{\pdiff{\polarFinal}{\ulocalInitial}} &
        \scalebox{\pdiffScaleFactor}{\pdiff{\polarFinal}{\vlocalInitial}} &
        \scalebox{\pdiffScaleFactor}{\pdiff{\polarFinal}{\azimuInitial}} &
        \scalebox{\pdiffScaleFactor}{\pdiff{\polarFinal}{\polarInitial}} &
        \scalebox{\pdiffScaleFactor}{\pdiff{\polarFinal}{\qopInitial}} \\ 
        \scalebox{\pdiffScaleFactor}{\pdiff{\qopFinal}{\ulocalInitial}} &
        \scalebox{\pdiffScaleFactor}{\pdiff{\qopFinal}{\vlocalInitial}} &
        \scalebox{\pdiffScaleFactor}{\pdiff{\qopFinal}{\azimuInitial}} &
        \scalebox{\pdiffScaleFactor}{\pdiff{\qopFinal}{\polarInitial}} &
        \scalebox{\pdiffScaleFactor}{\pdiff{\qopFinal}{\qopInitial}} \\[\lastRowMargin]  
    \end{pmatrix},    
\end{equation}
where the parameters at the initial and final frame are distinguished with a subscript of $\initial$ and $\final$, respectively.

The evaluation of $\jacobianFull$ can be facilitated by transforming the local track parameters to the global track parameters because their covariance matrices are easily transported during the track propagation and transformed back to the local track parameters after the propagation. Therefore, $\jacobianFull$ can be represented as the product of three sub-Jacobian matrices by applying the chain rule:  
\begin{equation}\label{eq:full_jacobi_decoupling}
    \jacobianFull = \pdiff{\localParsFinal}{\globalParsFinal} \pdiff{\globalParsFinal}{\globalParsInitial} \pdiff{\globalParsInitial}{\localParsInitial},   
\end{equation}
where $\globalParsInitial$ and $\globalParsFinal$ are the global track parameters at the initial and final frame, respectively. 

The coordinate transformation Jacobian matrices at the frame intersections, $\pdiff{\globalParsInitial}{\localParsInitial}$ and $\pdiff{\localParsFinal}{\globalParsFinal}$, are defined as $\jacobianLocalToGlobal$ and $\jacobianGlobalToLocal$, respectively:
\begin{align}\label{eq:transform_jacobi_general}
\jacobianLocalToGlobal & = \pdiff{\globalParsInitial}{\localParsInitial} = 
    \begin{pmatrix}[\matrixSpacingBig]
     \scalebox{\pdiffScaleFactor}{\pdiff{\posInitial}{\localInitial}} & \scalebox{\pdiffScaleFactor}{\pdiff{\posInitial}{\sphAngleInitial}} & 
     \scalebox{\pdiffScaleFactor}{\pdiff{\posInitial}{\qopInitial}} \\
     \scalebox{\pdiffScaleFactor}{\pdiff{\dirInitial}{\localInitial}} & \scalebox{\pdiffScaleFactor}{\pdiff{\dirInitial}{\sphAngleInitial}} & 
     \scalebox{\pdiffScaleFactor}{\pdiff{\dirInitial}{\qopInitial}} \\
     \scalebox{\pdiffScaleFactor}{\pdiff{\qopInitial}{\localInitial}} & 
     \scalebox{\pdiffScaleFactor}{\pdiff{\qopInitial}{\sphAngleInitial}} & 
     \scalebox{\pdiffScaleFactor}{\pdiff{\qopInitial}{\qopInitial}} \\[\lastRowMargin]
    \end{pmatrix},
    \nonumber \\
\jacobianGlobalToLocal & = \pdiff{\localParsFinal}{\globalParsFinal} =
    \begin{pmatrix}[\matrixSpacingBig]
     \scalebox{\pdiffScaleFactor}{\pdiff{\localFinal}{\posFinal}} & \scalebox{\pdiffScaleFactor}{\pdiff{\localFinal}{\dirFinal}} & 
     \scalebox{\pdiffScaleFactor}{\pdiff{\localFinal}{\qopFinal}} \\
     \scalebox{\pdiffScaleFactor}{\pdiff{\sphAngleFinal}{\posFinal}} & \scalebox{\pdiffScaleFactor}{\pdiff{\sphAngleFinal}{\dirFinal}} & 
     \scalebox{\pdiffScaleFactor}{\pdiff{\sphAngleFinal}{\qopFinal}} \\
     \scalebox{\pdiffScaleFactor}{\pdiff{\qopFinal}{\posFinal}} & \scalebox{\pdiffScaleFactor}{\pdiff{\qopFinal}{\dirFinal}} & \scalebox{\pdiffScaleFactor}{\pdiff{\qopFinal}{\qopFinal}} \\[\lastRowMargin]
    \end{pmatrix},
\end{align}
again, the parameters at the initial and final frame are distinguished with a subscript of $\initial$ and $\final$, respectively.

$\pdiff{\dirInitial}{\sphAngleInitial}$ and $\pdiff{\sphAngleFinal}{\dirFinal}$ are easily derived from \cref{eq:n_w_relations}:
\begin{align}\label{eq:n_w_derivatives}
\pdiff{\dirInitial}{\sphAngleInitial} & =
    \begin{pmatrix}[\matrixSpacingDefault]
    -\sin{\azimuInitial}\sin{\polarInitial}  & \cos{\azimuInitial}\cos{\polarInitial}  \\
    \cos{\azimuInitial}\sin{\polarInitial} & \sin{\azimuInitial}\cos{\polarInitial}  \\
     0 & -\sin{\polarInitial}
    \end{pmatrix}, \nonumber \\
\pdiff{\sphAngleFinal}{\dirFinal} & =
    \begin{pmatrix}[\matrixSpacingDefault]
     -\sin{\azimuFinal}/\sin{\polarFinal} & \cos{\azimuFinal}/\sin{\polarFinal} & 0 \\
     \cos{\azimuFinal}\cos{\polarFinal} & \sin{\azimuFinal}\cos{\polarFinal} & -\sin{\polarFinal} 
    \end{pmatrix}.
\end{align}
$\pdiff{\dirInitial}{\sphAngleInitial}$ is the right inverse of $\pdiff{\sphAngleFinal}{\dirFinal}$ in case the local-to-global and global-to-local coordinate transforms occur at the same frame without the track propagation. 

$\pdiff{\dirInitial}{\localInitial}$ and $\pdiff{\dirInitial}{\qopInitial}$ are zero matrices because $\sphAngleInitial$ is fixed by the definition of the partial derivative meaning that $\dirInitial$ is also fixed. The same logic can be applied to $\pdiff{\sphAngleFinal}{\posFinal}$ and $\pdiff{\sphAngleFinal}{\qopFinal}$ where $\dirFinal$ is fixed and so is $\sphAngleFinal$. Similarly, the derivatives of $\qopInitial$ and $\qopFinal$ with respect to other parameters are zero matrices because $\qop$ is fixed in the partial derivatives. The simplified representations of $\jacobianLocalToGlobal$ and $\jacobianGlobalToLocal$  with the zero and identity sub-matrices are given by:
\begin{align}\label{eq:transform_jacobi_simple}
\jacobianLocalToGlobal & =
    \begin{pmatrix}[\matrixSpacingBig]
     \scalebox{\pdiffScaleFactor}{\pdiff{\posInitial}{\localInitial}} & \scalebox{\pdiffScaleFactor}{\pdiff{\posInitial}{\sphAngleInitial}} & \scalebox{\pdiffScaleFactor}{\pdiff{\posInitial}{\qopInitial}} \\
     \zeroMatrix{3}{2} & \scalebox{\pdiffScaleFactor}{\pdiff{\dirInitial}{\sphAngleInitial}} & \zeroMatrix{3}{1} \\
     \zeroMatrix{1}{2} & \zeroMatrix{1}{2} & \identityMatrix{1}
    \end{pmatrix}, \nonumber \\
\jacobianGlobalToLocal & =    
    \begin{pmatrix}[\matrixSpacingBig]
     \scalebox{\pdiffScaleFactor}{\pdiff{\localFinal}{\posFinal}} & \scalebox{\pdiffScaleFactor}{\pdiff{\localFinal}{\dirFinal}} & \scalebox{\pdiffScaleFactor}{\pdiff{\localFinal}{\qopFinal}} \\
     \zeroMatrix{2}{3} & \scalebox{\pdiffScaleFactor}{\pdiff{\sphAngleFinal}{\dirFinal}} & \zeroMatrix{2}{1} \\
     \zeroMatrix{1}{3} & \zeroMatrix{1}{3} & \identityMatrix{1}
    \end{pmatrix}.
\end{align}

The transport Jacobian matrix in the global coordinate, $\pdiff{\globalParsFinal}{\globalParsInitial}$ of \cref{eq:full_jacobi_decoupling}, is defined as $\jacobianGlobalToGlobal$:

\begin{equation}\label{eq:transport_jacobi_def}
    \jacobianGlobalToGlobal = \pdiff{\globalParsFinal}{\globalParsInitial}.
\end{equation}
To find the general expression of $\jacobianGlobalToGlobal$, we can start by expanding the total differential $\dd{\globalParsFinal}$ using the fact that $\globalParsFinal$ is a function of $\globalParsInitial$ and $\pathlength$~\cite{WITTEK1980_Helix, WITTEK1981_Helix, STRANDLIE2006_Helix}:
\begin{equation}\label{eq:gf_variation}
   \dd{\globalParsFinal} = \left(\pdiff{\globalParsFinal}{\globalParsInitial}\right)_\pathlength \dd{\globalParsInitial} + \pdiff{\globalParsFinal}{\pathlength}\dd{\pathlength},
\end{equation}
where $\bigl(\pdiff{\globalParsFinal}{\globalParsInitial}\bigr)_\pathlength$ means $\pdiff{\globalParsFinal}{\globalParsInitial}$ with  fixed $\pathlength$. It should also be noted that $\globalParsFinal$ is constrained to intersect the final frame, represented as an implicit function ($\implicitF$):
\begin{equation}\label{eq:implicitF}
\implicitF(\globalParsFinal) = 0.
\end{equation}
The detailed equation of $\implicitF$ depends on the frame type. If $\implicitF$ is differentiable with respect to $\globalParsFinal$, the following  equation holds:
\begin{equation}
    \pdiff{\implicitF}{\globalParsFinal} \cdot \dd{\globalParsFinal} = 0,
\end{equation}
which can be utilized to obtain the expression of $\dd{\pathlength}$ by taking an inner product between $\pdiff{\implicitF}{\globalParsFinal}$ and $\dd{\globalParsFinal}$ of \cref{eq:gf_variation}:
\begin{equation}
    \dd{\pathlength} = -\left(\pdiff{\implicitF}{\globalParsFinal} \cdot \pdiff{\globalParsFinal}{\pathlength}\right)^{-1} \left[ \pdiff{\implicitF}{\globalParsFinal} \left(\pdiff{\globalParsFinal}{\globalParsInitial}\right)_\pathlength \right] \cdot \dd{\globalParsInitial}.
\end{equation}
The above equation can be fed back into \cref{eq:gf_variation} to absorb the  $\dd{\pathlength}$ term and to obtain $\jacobianGlobalToGlobal$ defined in \cref{eq:transport_jacobi_def}:
\begin{equation}\label{eq:jac_g2g}
    \jacobianGlobalToGlobal = \left[ \identityMatrix{7} - \left(\pdiff{\implicitF}{\globalParsFinal} \cdot \pdiff{\globalParsFinal}{\pathlength}\right)^{-1} \pdiff{\globalParsFinal}{\pathlength} \pdiff{\implicitF}{\globalParsFinal} \right] \left( \pdiff{\globalParsFinal}{\globalParsInitial} \right)_\pathlength.
\end{equation}
The $\pdiff{\globalParsFinal}{\pathlength}$ component of $\jacobianGlobalToGlobal$ has an analytical solution:
\begin{equation}\label{eq:dgds}
    \transpose{\left(\pdiff{\globalParsFinal}{\pathlength}\right)} = \begin{pmatrix}
    \transpose{\dirFinal} & \qopFinal \transpose{(\dirFinal \times \bfieldFinal)} & \frac{\qopFinal^{3} \energyFinal}{\charge^2} \left( -\pdiff{\energyFinal}{\pathlength} \right)
    \end{pmatrix},
\end{equation}
where $\bfieldFinal$ and $\energyFinal$ are the magnetic field and the energy ($\energy$) of the charged particle at $\globalParsFinal$, respectively. $\left(\stoppingPower\right)$ is the stopping power~\cite{WORKMAN2022_PDG} of the material at a point where the particle is propagating.

The $\bigl(\pdiff{\globalParsFinal}{\globalParsInitial}\bigr)_\pathlength$ can be either analytically calculated for helical tracks with homogeneous magnetic fields or numerically calculated for tracks with inhomogeneous magnetic fields and materials. The helix model derived in~\cite{WITTEK1980_Helix, WITTEK1981_Helix, STRANDLIE2006_Helix} is introduced in \ref{appendix:helix}. The fourth order RKN method is one of the most widely used numerical integration methods in the high energy physics experiments. In \ref{appendix:rkn4}, we also introduce its calculation of $\bigl(\pdiff{\globalParsFinal}{\globalParsInitial}\bigr)_\pathlength$  which is derived in~\cite{BUGGE1981_RK, LUND2009_CovarianceTransport}.

In the following subsections, the remaining block matrices of the coordinate transform Jacobian matrices and $\pdiff{\implicitF}{\globalParsFinal}$ of $\jacobianGlobalToGlobal$, which are specific to the frame type, will be derived for the bound and perigee frames.

\subsection{Bound frame}\label{subsec:bound_frame}

\begin{figure}[ht]
    \begin{center}
        \includegraphics[width=\frameFigureSize\linewidth]{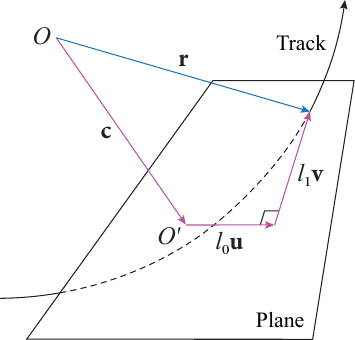}
    \end{center}        
    \caption{Track parametrization for the bound frame where $O$ and $O'$ represent the origins of the global and local coordinate systems, respectively.}
    \label{fig:bound_frame}
\end{figure}

In many tracking detectors, the local track parameters need to be described on planar surfaces. The local coordinate frame used to describe such planar surfaces is referred to a bound frame whose $\ubasis$ and $\vbasis$ are defined on the plane. The track representation of \cref{eq:r_in_uv} on the bound frame is illustrated in \cref{fig:bound_frame}. The condition for the frame intersection is met when the vector from the origin of the local coordinate system to the track position, i.e., ($\pos - \frameCenter$),  is on the plane and perpendicular to the plane normal vector ($\planeNormal$), which is defined as $\ubasis \times \vbasis$. The condition can be represented as the implicit function of \cref{eq:gf_variation}:
\begin{equation}\label{eq:bound_implicit}
    \boundImplicit(\globalPars) = \planeNormal \cdot (\pos - \frameCenter) = 0.
\end{equation}
The origin of the local coordinate system can be any point on the plane, which will satisfy the implicit condition of \cref{eq:bound_implicit}, and $(\ulocal, \vlocal)$ of $\local$ and $\frameCenter$ will be set accordingly.

For the coordinate transformation Jacobian matrices of \cref{eq:transform_jacobi_simple}, $\pdiff{\posInitial}{\localInitial}$ and $\pdiff{\localFinal}{\posFinal}$  can be obtained from \cref{eq:mu_in_uv,eq:r_in_uv} using the fact that $\ubasis$, $\vbasis$, and $\frameCenter$ are independent of $\pos$ and $\local$:
\begin{equation}\label{eq:dr_dl}
\pdiff{\posInitial}{\localInitial}=
    \begin{pmatrix}
     \ubasisInitial & \vbasisInitial 
    \end{pmatrix},
\quad
\pdiff{\localFinal}{\posFinal}=    
    \begin{pmatrix}[\matrixSpacingDefault]
     \transpose{\ubasisFinal} \\
     \transpose{\vbasisFinal}
    \end{pmatrix}.
\end{equation} 
It is easy to confirm that $\pdiff{\posInitial}{\localInitial}$ is the right inverse of $\pdiff{\localFinal}{\posFinal}$ in case the coordinate transforms occur at the same position of the same frame.

The remaining off-diagonal sub-matrices are zero matrices because $\pos$ of \cref{eq:r_in_uv} and $\local$ of \cref{eq:mu_in_uv} of the bound frame are independent of $\sphAngle$, $\dir$ and $\qop$. Thus, $\jacobianLocalToGlobal$ and $\jacobianGlobalToLocal$ are simplified into block diagonal matrices:
\begin{align}\label{eq:transform_jacobi_bound}
\jacobianLocalToGlobal=
    \begin{pmatrix}[\matrixSpacingBig]
     \scalebox{\pdiffScaleFactor}{\pdiff{\posInitial}{\localInitial}} & \zeroMatrix{3}{2} & \zeroMatrix{3}{1} \\
     \zeroMatrix{3}{2}  & \scalebox{\pdiffScaleFactor}{\pdiff{\dirInitial}{\sphAngleInitial}} & \zeroMatrix{3}{1} \\
     \zeroMatrix{1}{2} & \zeroMatrix{1}{2} & \identityMatrix{1}
    \end{pmatrix}, \nonumber \\
\jacobianGlobalToLocal=    
    \begin{pmatrix}[\matrixSpacingBig]
     \scalebox{\pdiffScaleFactor}{\pdiff{\localFinal}{\posFinal}} & \zeroMatrix{2}{3} & \zeroMatrix{2}{1}  \\
     \zeroMatrix{2}{3} & \scalebox{\pdiffScaleFactor}{\pdiff{\sphAngleFinal}{\dirFinal}} & \zeroMatrix{2}{1} \\
     \zeroMatrix{1}{3} & \zeroMatrix{1}{3} & \identityMatrix{1} 
    \end{pmatrix}.
\end{align}
It is straightforward to confirm that 
$\jacobianLocalToGlobal$ is the right inverse of $\jacobianGlobalToLocal$, which preserves the covariance matrix after for the local-to-global-to-local transformation on the same frame.

In order to calculate $\jacobianGlobalToGlobal$ for the bound frames, it is sufficient to obtain $\pdiff{\implicitF}{\globalParsFinal}$ from \cref{eq:bound_implicit}:
\begin{equation}
    \pdiff{\implicitF}{\globalParsFinal} =     
    \begin{pmatrix}
        \transpose{\planeNormalFinal} & \zeroMatrix{1}{4}
    \end{pmatrix},
\end{equation}
where $\planeNormalFinal$ is the plane normal vector of the final frame. The above equation can be put into \cref{eq:jac_g2g} with \cref{eq:dgds}:
\begin{equation}\label{eq:jac_g2g_bound}
    \jacobianGlobalToGlobal = \left[ \identityMatrix{7} - \frac{1}{\planeNormalFinal \cdot \dirFinal} \pdiff{\globalParsFinal}{\pathlength}
    \begin{pmatrix}
        \transpose{\planeNormalFinal} & \zeroMatrix{1}{4}
    \end{pmatrix} \right] \left(\pdiff{\globalParsFinal}{\globalParsInitial}\right)_\pathlength.
\end{equation}
$\jacobianGlobalToGlobal$ diverges when the track direction is almost parallel to the plane at the final frame intersection, where even a very small variation of track parameter at the initial frame can affect a large offset to the intersection points at the final frame.

\subsection{Perigee frame}\label{subsec:perigee_frame}

\begin{figure}[ht]
    \begin{center}
        \includegraphics[width=\frameFigureSize\linewidth]{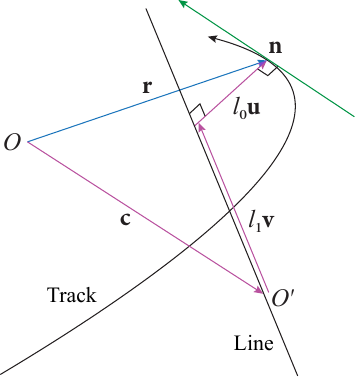}
    \end{center}        
    \caption{Track parametrization for the perigee frame where $O$ and $O'$ represent the origins of the global and local coordinate systems, respectively.}
    \label{fig:perigee_frame}
\end{figure}

The perigee frame~\cite{BILLOIR1992_Perigee} provides a track parametrization to describe the distance between a track and a straight line. The track representation of the perigee frame based on \cref{eq:r_in_uv} is illustrated in \cref{fig:perigee_frame}. Here $\vbasis$ is a unit vector parallel to the line direction and $\ubasis$ is perpendicular to $\dir$ such that $\ulocal$ is the distance between the track and the line. This track parametrization is useful for the vertex reconstruction and wire measurements of straw trackers and drift chambers where the closest approach between the track and a straight line needs to be found. The closest approach is met when $\dir$ is orthogonal to $\ubasis$, therefore, the implicit function of the frame intersection can be defined as follows:
\begin{equation}\label{eq:perigee_implicit}
 \perigeeImplicit(\globalPars) = \dir \cdot \ulocal \ubasis= \dir \cdot \{ (\pos - \frameCenter) -  [(\pos - \frameCenter) \cdot \vbasis] \vbasis \} = 0,
\end{equation}
where \cref{eq:r_in_uv,eq:mu_in_uv} are used to represent $\ulocal \ubasis$ as a function of $\pos$. The origin of the local coordinate system must lie along the line to maintain orthogonality between $\dir$ and $\ubasis$. This one-dimensional translation of the origin of the local coordinate system along the line sets $\vlocal$ and $\frameCenter$ accordingly while $\ulocal$ remains invariant. $\ubasis$ is defined as follows to satisfy the condition of \cref{eq:perigee_implicit}:
\begin{equation}\label{eq:perigee_ubasis_def}
 \ubasis = \frac{\vbasis \times \dir}{| \vbasis \times \dir|},
\end{equation}
where the sign of $\ulocal$ follows the sign of $\ubasis\cdot(\pos-\frameCenter)$. There are a few differences in the perigee frame definition between this paper and \cite{BILLOIR1992_Perigee}, such as the sign convention and a constant multiplier, but the basic principle remains the same.

For the coordinate transformation Jacobian matrices, \cref{eq:dr_dl} holds as for the bound frame. $\pdiff{\posInitial}{\qopInitial}$ and $\pdiff{\localFinal}{\qopFinal}$ are zero matrices because $\pos$ of \cref{eq:r_in_uv} and $\local$ of \cref{eq:mu_in_uv}  are not functions of $\qop$. That $\pdiff{\localFinal}{\dirFinal}$ is a zero matrix can be deduced heuristically (see \ref{appendix:perigee_proof} for the detailed proof): As $\posFinal$ is fixed, the variation of $\dirFinal$ can be understood as a rotation around $\ubasisFinal$ to preserve $\posFinal$. However, $\pdiff{\posInitial}{\sphAngleInitial}$ is a nonzero matrix because the intersection point can rotate around $\vbasisInitial$ while conserving the elements of $\localInitial$. In summary, the coordinate transform Jacobian matrices can be written as the following:

\begin{align}\label{eq:transform_jacobi_perigee}
\jacobianLocalToGlobal & =
    \begin{pmatrix}[\matrixSpacingBig]
     \scalebox{\pdiffScaleFactor}{\pdiff{\posInitial}{\localInitial}} &  \scalebox{\pdiffScaleFactor}{\pdiff{\posInitial}{\sphAngleInitial}} & \zeroMatrix{3}{1} \\
     \zeroMatrix{3}{2} & \scalebox{\pdiffScaleFactor}{\pdiff{\dirInitial}{\sphAngleInitial}} & \zeroMatrix{3}{1} \\
     \zeroMatrix{1}{2} & \zeroMatrix{1}{2} & \identityMatrix{1}
    \end{pmatrix}, \nonumber \\
\jacobianGlobalToLocal & =    
    \begin{pmatrix}[\matrixSpacingBig]
     \scalebox{\pdiffScaleFactor}{\pdiff{\localFinal}{\posFinal}} & \zeroMatrix{2}{3} & \zeroMatrix{2}{1} \\
     \zeroMatrix{2}{3} & \scalebox{\pdiffScaleFactor}{\pdiff{\sphAngleFinal}{\dirFinal}} & \zeroMatrix{2}{1} \\
     \zeroMatrix{1}{3} & \zeroMatrix{1}{3} & \identityMatrix{1} 
    \end{pmatrix}.
\end{align} 
Here, we only present the final form of $\pdiff{\posInitial}{\azimuInitial}$ for $\pdiff{\posInitial}{\sphAngleInitial}$ (see \ref{appendix:perigee_proof} for the detailed derivation):
\begin{equation}\label{eq:drdphi}
    \drdphiInitial.
\end{equation}
$\pdiff{\posInitial}{\polarInitial}$ follows the same equation by replacing $\azimuInitial$ with $\polarInitial$. Like the bound frame, the product of $\jacobianGlobalToLocal$ and $\jacobianLocalToGlobal$ on the same perigee frame is the identity matrix as $\pdiff{\localFinal}{\posFinal} \pdiff{\posInitial}{\sphAngleInitial}$ is the zero matrix due to the orthogonality between \cref{eq:drdphi} and \cref{eq:dr_dl}.

\begin{figure*}[ht]
    \centering
    \includegraphics[width=\geometryFigureSize\linewidth]{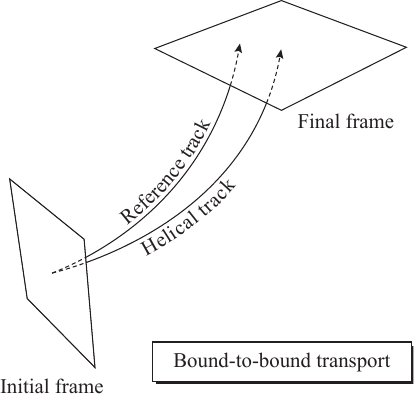}
    \quad \quad \quad \quad 
    \includegraphics[width=\geometryFigureSize\linewidth]{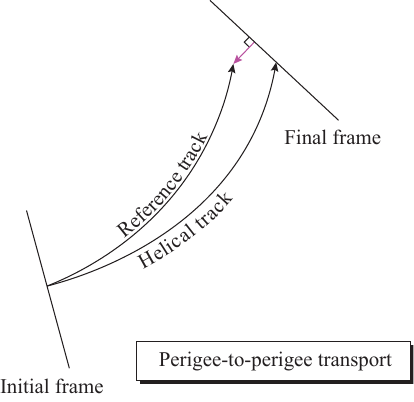}
    \caption{Illustrations of (left) the geometry for bound-to-bound transport and (right) the geometry for perigee-to-perigee transport. The helical track starts from the the local origin of the initial frame, which is set to the global coordinate origin, to determine that of the final frame based on randomly sampled momentum vector and propagation length. The reference track of the same initial track parameters is propagated to the final frame for the estimation of $\jacobianFull$. Under the influence of an inhomogeneous magnetic field and a material, the final frame intersection of the reference track is deviated from the local origin. The magenta arrow on the final perigee frame represents the segment of the closest approach, i.e., $\ulocalFinal\ubasisFinal$.}
    \label{fig:geometry}
\end{figure*}

For the derivation of $\jacobianGlobalToGlobal$, we can take the same approach as for the bound frame case and obtain $\pdiff{\implicitF}{\globalParsFinal}$ from \cref{eq:perigee_implicit}:
\begin{equation}
    \pdiff{\implicitF}{\globalParsFinal} =
    \begin{pmatrix}
        \transpose{[\dirFinal - (\dirFinal \cdot \vbasisFinal)\vbasisFinal]} & \transpose{\ulocalFinal \ubasisFinal} & \zeroMatrix{1}{1}
    \end{pmatrix},
\end{equation}
where $\ubasisFinal$ and $\vbasisFinal$ are the orthonormal basis of the final frame. Substituting into \cref{eq:jac_g2g} and using \cref{eq:dgds}:
\begin{align}
    \jacobianGlobalToGlobal & = \Biggl[ \identityMatrix{7} - \frac{1}{1- (\dirFinal \cdot \vbasisFinal)^2 + \perigeeAdditionalTerm} \nonumber \\
    & \qquad \quad \times \pdiff{\globalParsFinal}{\pathlength}  \begin{pmatrix}
        \transpose{[\dirFinal - (\dirFinal \cdot \vbasisFinal)\vbasisFinal]} & \transpose{\ulocalFinal \ubasisFinal} & \zeroMatrix{1}{1}
    \end{pmatrix}\Biggr] \left(\pdiff{\globalParsFinal}{\globalParsInitial}\right)_\pathlength,
\end{align}
where $\perigeeAdditionalTerm = \ulocalFinal \ubasisFinal \cdot \qopFinal (\dirFinal \times \bfieldFinal)$.

The exact divergence condition of $\jacobianGlobalToGlobal$ is complicated, but can be simplified for tracks with small $\perigeeAdditionalTerm$, i.e. on the scale of ${\ulocal\qop|\bfield|}$. This simplification has widespread validity because  $\ulocal$ is limited in a range much smaller than $\qop |\bfield|$ for most experimental setups. For example, a particle with the momentum of \SI[per-mode=symbol]{1}{\giga\eV\per\clight} in a \SI{1}{\tesla} longitudinal magnetic field has $\qop \abs{\bfield}$ = \SI{0.003}{\per\centi\meter} in the $xy$-plane, and the values of $\ulocal$ for wire measurements are usually in the scale of \SI{1}{\centi\meter}, such that ${\ulocal\qop |\bfield|}$ becomes negligible. Under this assumption, the divergence condition is met when the track is almost parallel to $\vbasis$, for the same reason explained in \cref{subsec:bound_frame}.

\section{Validation}\label{sec:validation}

We validate the derivation of the Jacobian matrices for bound and perigee frames---following the validation procedure presented in \cite{LUND2009_CovarianceTransport}---by simulating muon tracks propagating from an initial frame to a final frame in the presence of magnetic fields and detector material. The default configuration has an inhomogeneous magnetic field of the Open Data Detector (ODD)~\cite{GESSINGER2023_ODDACAT, CORENTIN2022_ODD} and a uniform material distribution of cesium iodide (CsI). The ODD magnetic field vectors are known at intervals of \SI{100}{\mm} in the $x$, $y$, and $z$ directions. Within a cube with an edge of \SI{1}{\meter} and center at the global origin, the field strength ranges from \SI{1.979}{\tesla} to \SI{2.014}{\tesla}, and the transverse magnetic field is almost negligible with magnitude less than 0.1\% of the longitudinal magnetic field strength. The field gradually increases closer to the $z$-axis and further from $xy$-plane at $z=0$.  CsI, consisting of 51.2\% cesium and 48.8 \% iodide, has the density of \SI[per-mode=symbol]{4.51}{\gram\per\cm^3}~\cite{GROOM2021_NUCLEARTABLE}, which is dense enough to observe the effect of energy loss clearly. For the propagation through the inhomogeneous magnetic field, field vectors at arbitrary points are linearly interpolated~\cite{SWATMAN2023_covfie}. To simulate material interactions, the mean energy loss from inelastic collisions with the electrons of atoms, which follows the Bethe equation~\cite{BETHE1930_Bethe}, is considered as a muon stopping power and the equations used are described in \ref{appendix:mean_eloss}. Radiative energy losses such as Bremsstrahlung, which are negligible for the muon momenta of interest, are neglected and random fluctuations from the multiple scattering, which can not be represented as a Jacobian matrix, are not considered.

The muon tracks propagated for the estimation of $\jacobianFull$ will be called \textit{reference tracks} whose $\localParsInitialReference$ and $\covarianceInitialReference$ are randomly sampled. The reference tracks always start at the local origin of the initial frame and  $\azimu$, $\polar$ and $\momentum$ are randomly sampled from  uniform distributions in the range of [$-\pi$, $\pi$], [0, $\pi$] and  [\SI[per-mode=symbol]{500}{\mega\eV\per\clight}, \SI[per-mode=symbol]{100}{\giga\eV\per\clight}], respectively. In the momentum range used, the mean energy loss per distance monotonically increases from \SI[per-mode=symbol]{0.57}{\mega\eV\per\mm} to \SI[per-mode=symbol]{0.88}{\mega\eV\per\mm} as a function of momentum. For the muon momenta large enough, the total energy loss can be approximated as a product of the mean energy loss per distance and the actual track propagation length.

Every reference track is propagated in its own geometries consisting of two geometrical objects which are defined for the initial and final frames. Tests are performed separately for a geometry with two planes with bound frames and a second geometry with two lines with perigee frames, as illustrated in \cref{fig:geometry}. The $\jacobianFull$ is evaluated for track propagation between two bound frames (\textit{bound-to-bound transport}) and between two perigee frames (\textit{perigee-to-perigee transport}) \footnote{While the jacobian evaluation with a bound-to-perigee transport is skipped due to redundancy with the other two cases, it may also be useful in the vertex reconstruction of silicon tracker experiments in case of propagating a track on a planar surface to the beam line backwardly.}. The local origin of the initial frame is set to the origin of the global coordinate system. The local origin of the final frame is set by propagating a helical track which shares the initial track parameters of the reference track. The helical track assumes a homogeneous magnetic field parallel to the $z$-axis with a strength of \SI{1.996}{\tesla}, which is the average value of the ODD magnetic field. The helical path length from the initial to the final frame is randomly selected from a uniform distribution with a range of [\SI{50}{\mm}, \SI{500}{\mm}]. The $\planeNormal$ of the bound frame is set to the direction of the helical track ($\helicalDir$) at the origin of the local coordinate system, and $\ubasis$ is set to the curvilinear vector of $\helicalDir$, namely $\curvilinearVector = \frac{\unitZ \times \helicalDir}{|\unitZ \times \helicalDir|}$, where $\unitZ$ is the unit vector along the $z$-axis.  The $\vbasis$ of the perigee frame is the clockwise $\nicefrac{\pi}{2}$-rotation of $\helicalDir$ around $\curvilinearVector$. This makes $\vbasis$ of the perigee frames identical to that of the bound frames.

For generality, we rotate the final frames with random Euler angles in the order of local $z$-$x$-$z$ axes where the local $x$-axis is $\ubasis$ and the local $z$-axis is $\helicalDir$. The first rotation around the local $z$-axis is randomly sampled from a uniform distribution with a range of [0, $2\pi$]. The second rotation around the new local $x$-axis is performed by randomly sampling the cosine of the angle from a uniform distribution with a range of [$\nicefrac{1}{\sqrt{2}}$, 1], corresponding to the rotation angle in a range of [0, $\nicefrac{\pi}{4}$], and the sign of the rotation angle is also sampled at random. The final rotation around the new local $z$-axis is randomly sampled from a uniform distribution with a range of [0, $2\pi$]. The initial frames are rotated only once around the local $z$-axis in the range of [0, $2\pi$]. The size of the final frames, corresponding to the maximum limits of $\ulocalFinal$ and $\vlocalFinal$, are set 1.5 times larger than the helical path length which is enough to ensure that all tracks intersect the frames.

\begin{figure}[ht]
    \begin{center}
    \includegraphics[width=\rkStepFigureSize\linewidth]{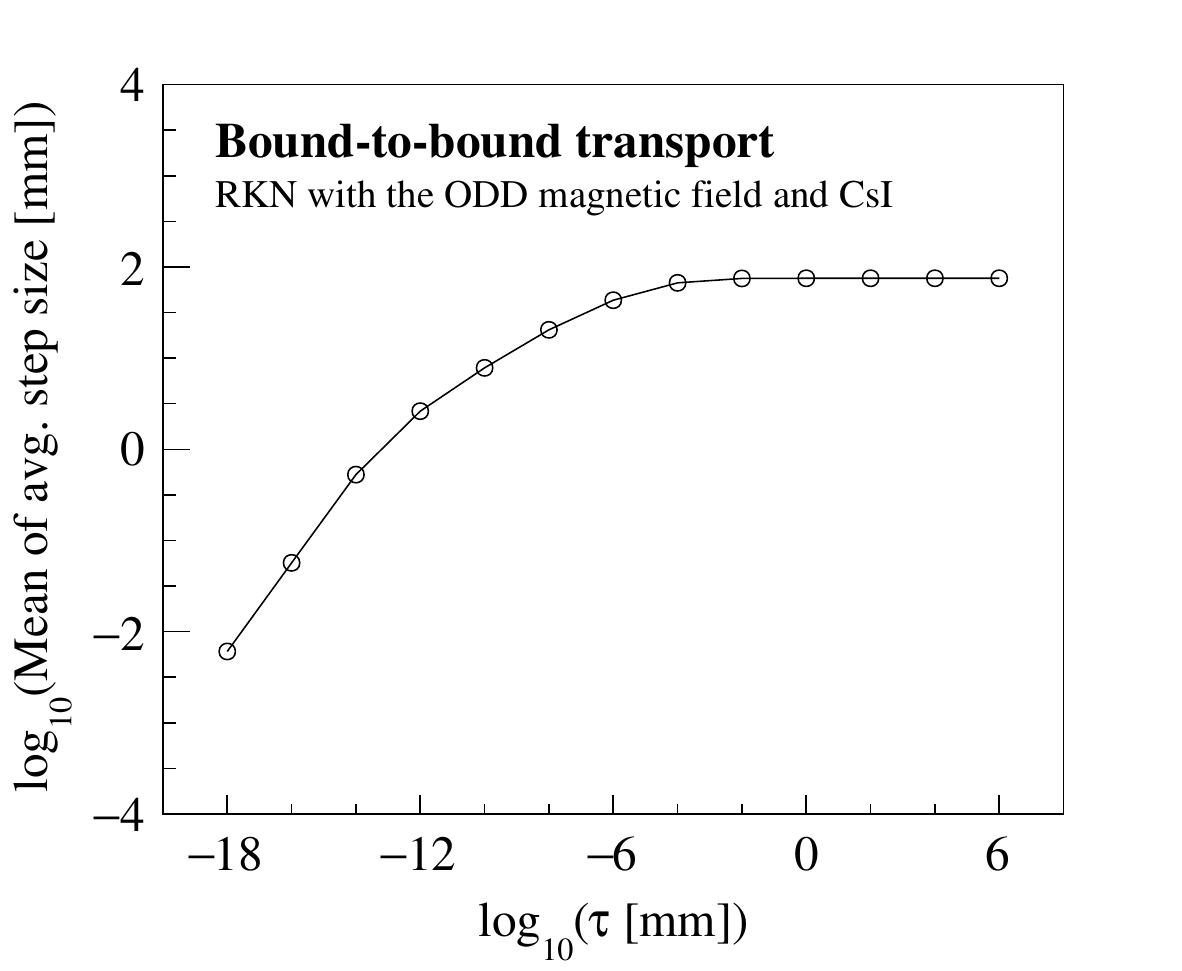}
    \caption{Mean value of the step sizes averaged over $\nTracks$ reference tracks as a function of the error tolerance $\globalError$. The reference tracks are propagated for the bound-to-bound transport with the ODD magnetic field and CsI.}
    \label{fig:mean_step_size}
    \end{center}
\end{figure}

We use the adaptive fourth order RKN method~\cite{LUND2009_CovarianceTransport, LUND2009_RKNPropagation} as a numerical integration model for the propagation of muon tracks in magnetic fields and materials. In the adaptive fourth order RKN method, the step size is decreased if the estimated local truncation error of the current step is larger than an \textit{error tolerance} ($\globalError$), and increased to save the computation  if the local error is smaller than $\globalError$. The local truncation error of the step is estimated by measuring the distance between $\pos$ of the fourth and third order solutions of the current step (see \ref{appendix:rkn4})~\cite{LUND2009_RKNPropagation, PRESS2007_Recipe}. The mean values of the step sizes are measured as a function of $\globalError$ with the geometry of the bound frames as shown in \cref{fig:mean_step_size}. In addition to the adaptive step size scaling of the RKN method, the step size is limited by the distance to the expected intersection with the frame which is estimated for every step using a straight, tangential ray. The same procedure is repeated until the distance to the frame intersection is less than \SI{e-9}{\mm}. These processes for the local-to-local frame transport are summarized as a general function $\InitialToFinalMapping$:
 
\begin{equation}\label{eq:parameter_mapping}
    \localParsFinalReference = \InitialToFinalMapping(\localParsInitialReference).
\end{equation}
The track propagation represented by \cref{eq:parameter_mapping} is executed by the detray library~\cite{SALZBURGER2023_detray} and all calculations are performed with 64-bit floating point numbers to maximize precision.

In \cref{subsec:numerical_diff}, we compare the difference between $\jacobianFull$ of the reference track and the result of numerical differentiation. We study the difference as a function of $\globalError$ to understand its impact on the precision. We repeat the study with the various configurations to investigate the robustness of the Jacobian matrix evaluation against the computation complexity increased by the gradients of magnetic fields or the presence of materials.

In \cref{subsec:cov_transport}, the validation study is extended to the covariance matrix transport because the correctness of Jacobian matrices is not sufficient to justify the first-order Taylor expansion in \cref{eq:covariance_transport}. We attempt to verify that $\covarianceFinal$ of the reference tracks agrees with the track parameter distributions at the final frame by transporting tracks whose initial parameters are smeared by $\covarianceInitial$.

\begin{figure*}[ht]
    \begin{center}
    \begin{subfigure}{\rkToleranceFigureSize\textwidth}
    \includegraphics[width=1.0\linewidth]{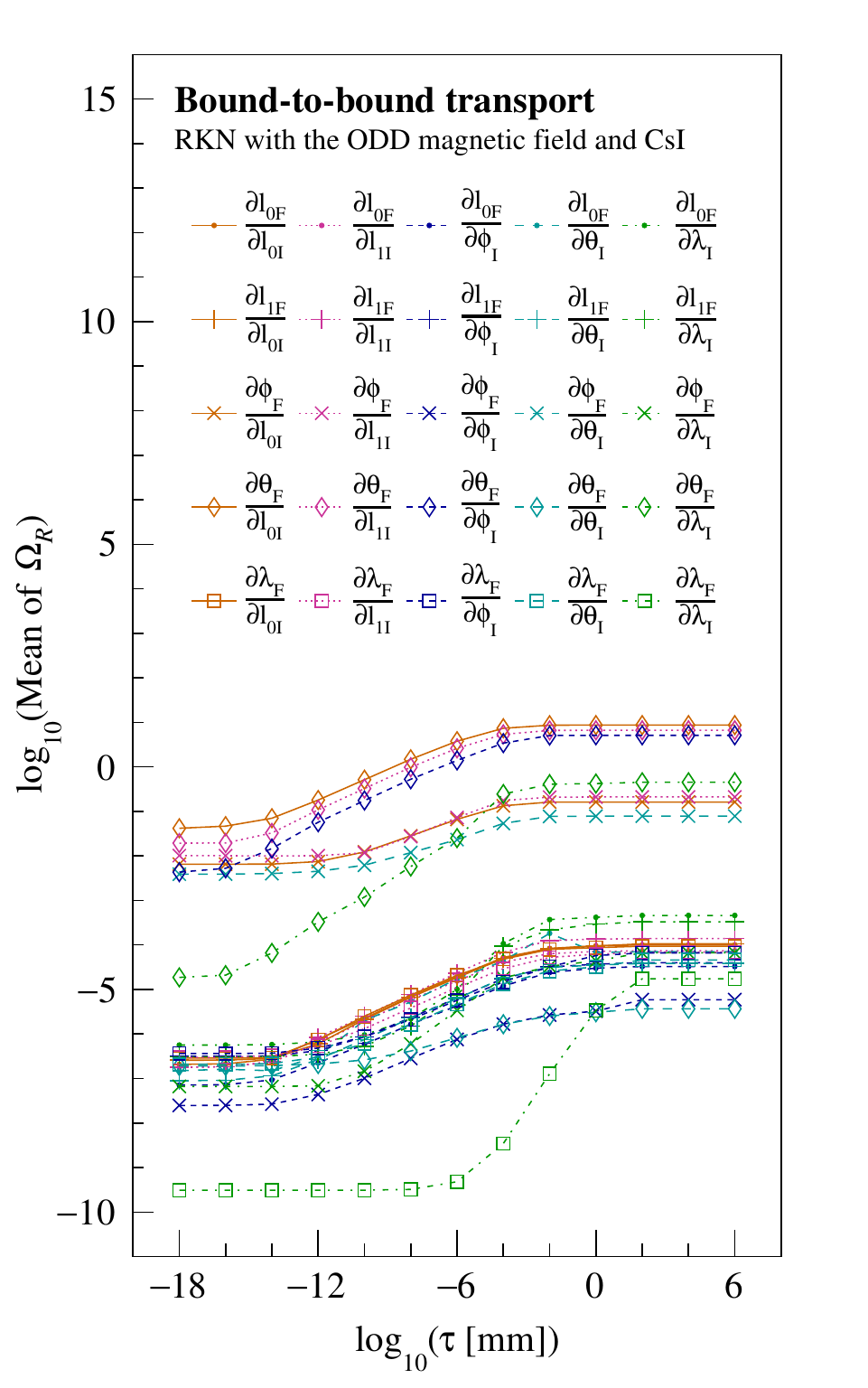}
    \end{subfigure}
    \begin{subfigure}{\rkToleranceFigureSize\textwidth}
    \includegraphics[width=1.0\linewidth]{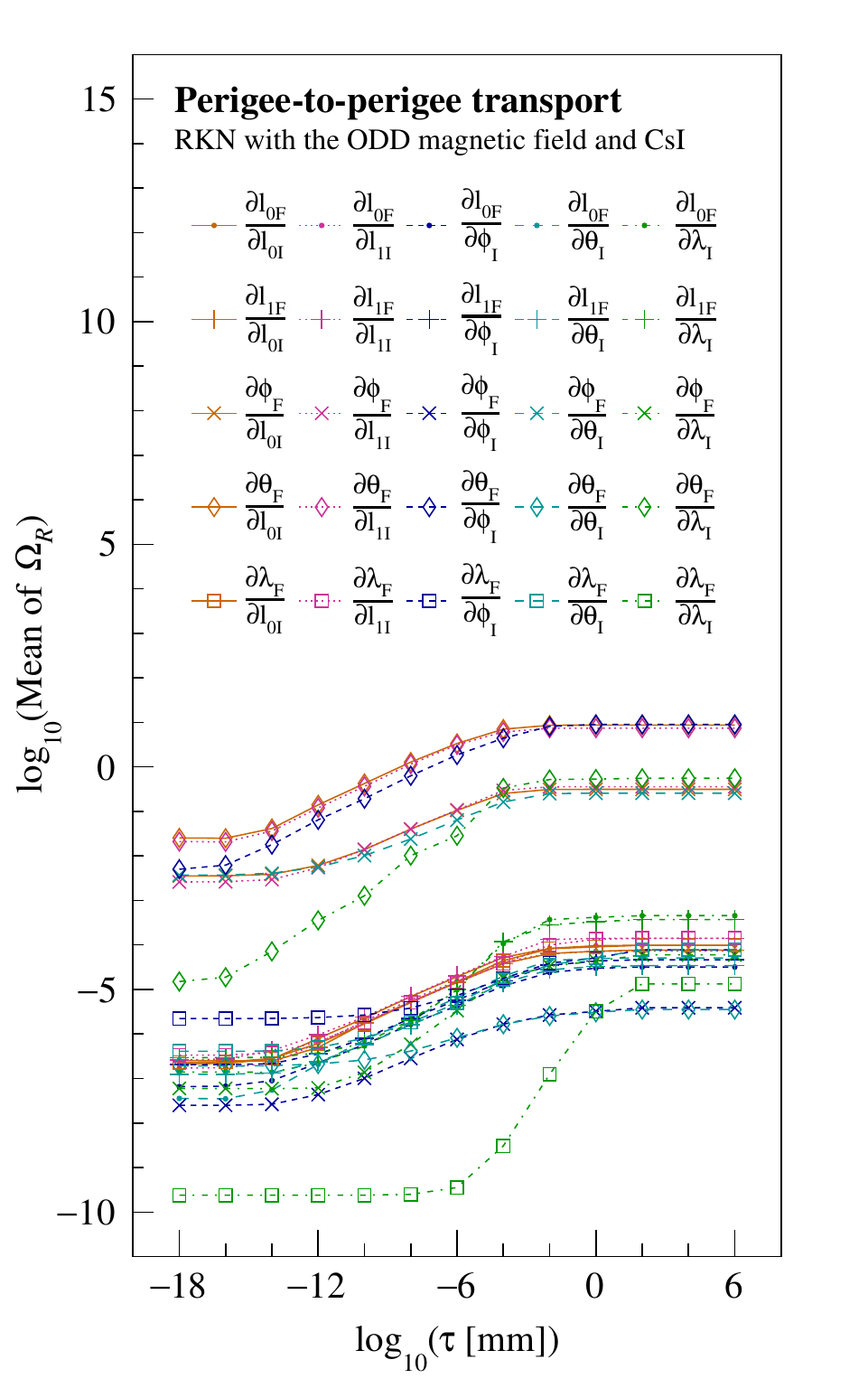}
    \end{subfigure}
    \caption{Mean value of $\absRelResidual$ for each of $\jacobianFull$ elements from (left) bound-to-bound transport and (right) perigee-to-perigee transport with the ODD magnetic field and CsI. The $\absRelResidual$ is obtained as a function of $\globalError$ of the reference tracks when $\globalError$ of the shifted tracks is set to $\toleranceShitedTracks$~\SI{}{\mm}.} 
    \label{fig:residuals_full_setup}
    \end{center}        
\end{figure*}

\begin{figure}[ht]
    \begin{center}
    \includegraphics[width=\histogramFigureSize\linewidth]{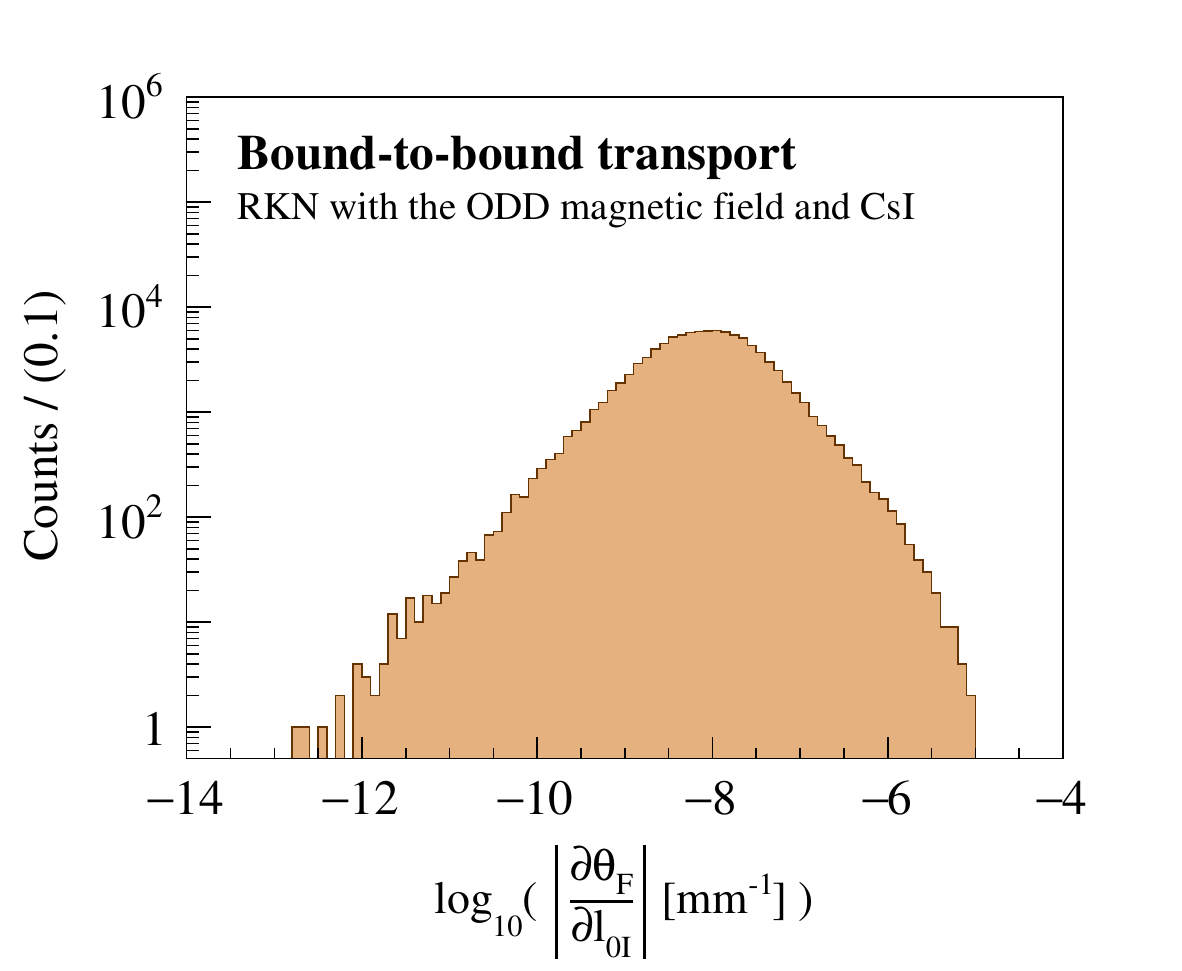}
    \includegraphics[width=\histogramFigureSize\linewidth]{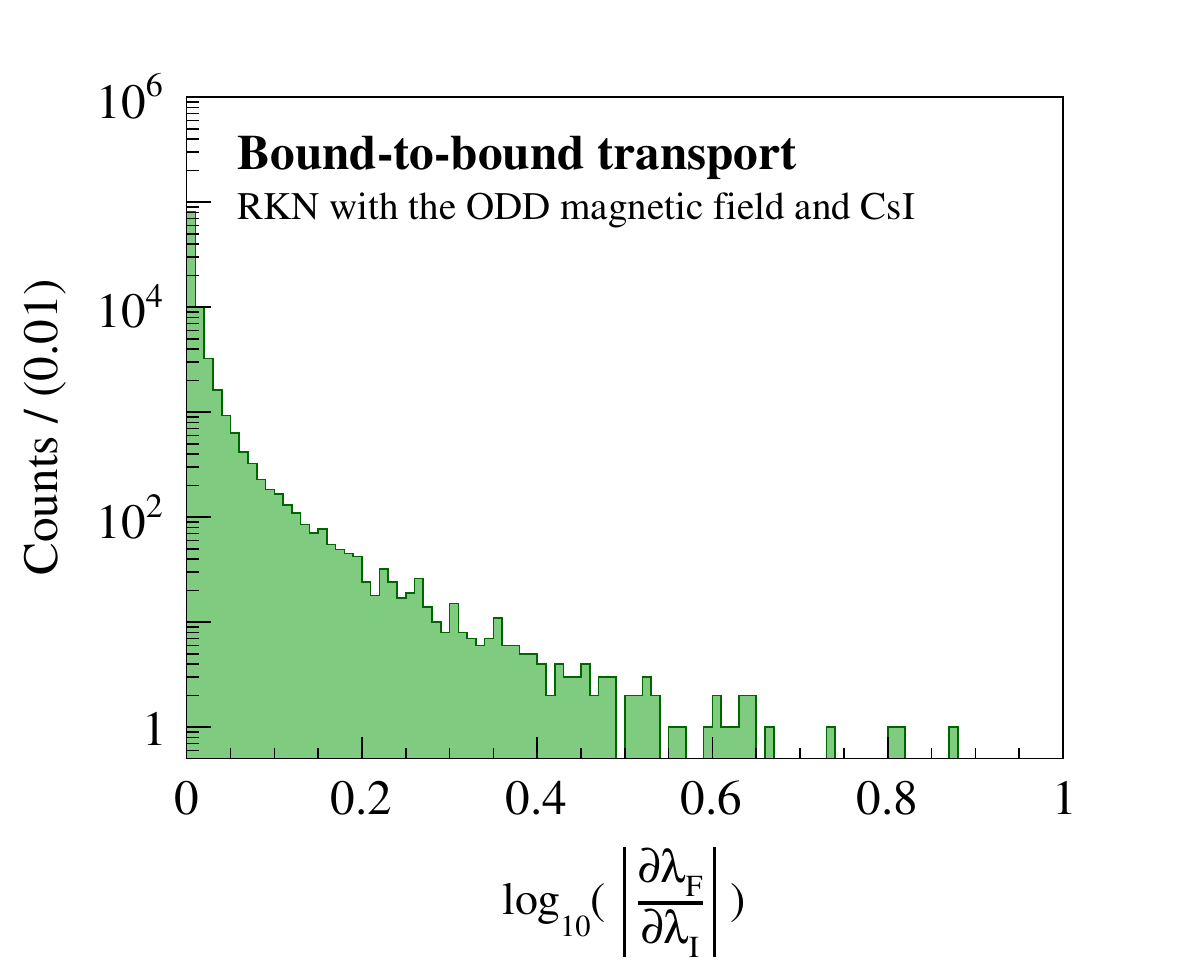}
    \caption{Histograms of (top) $\abs{\pdiff{\polarFinal}{\ulocalInitial}}$ and (bottom) $\abs{\pdiff{\qopFinal}{\qopInitial}}$ of $\jacobianFull$ from the bound-to-bound transport with the ODD magnetic field and CsI. The $\globalError$ of the reference tracks is set to $\toleranceMin$~\SI{}{\mm}.}
    \label{fig:residual histogram}
    \end{center}
\end{figure}

\subsection{Validation of $\jacobianFull$ using numerical differentiation}\label{subsec:numerical_diff}

The numerical differentiation refers to algorithms that numerically calculate the derivative of a function by estimating a change of the function output with respect to a change ($\dpl$) of the function input. For the numerical differentiation of \cref{eq:parameter_mapping}, we apply changes of $\pm \dpl_\indexJ$ to $\indexJ$-th elements of $\localParsInitialReference$ to generate two \textit{shifted tracks}. The error propagation of the shifted tracks can be skipped because we are only interested in the local track parameters at the final frame, namely $\localParsFinalPlus^{\indexJ}$ and $\localParsFinalMinus^{\indexJ}$: 
\begingroup
\addtolength{\jot}{\equationSpacingA}
\begin{align}
    \localParsFinalPlus^{\indexJ} & = \InitialToFinalMapping(\localParsInitial + \dpl_\indexJ \stdbasis_{\indexJ}), \nonumber \\
    \localParsFinalMinus^{\indexJ} & = \InitialToFinalMapping(\localParsInitial - \dpl_\indexJ \stdbasis_{\indexJ}),
\end{align}
\endgroup
where $\stdbasis_{\indexJ}$ is the standard basis where the element at the $\indexJ$-th index is one and the other four are zero. We can numerically estimate the element of $\jacobianFull$ by calculating a symmetric difference quotient:

\begin{equation}\label{eq:num_diff}
    \jacobianFull_{\leftBraIdx\indexI,\indexJ\rightBraIdx} =  \frac{\localPars^{\indexJ}_{\final+\leftBraIdx\indexI\rightBraIdx} - \localPars^{\indexJ}_{\final-\leftBraIdx\indexI\rightBraIdx}}{2 \dpl_{\indexJ}} + \mathrm{T.E.} = \symmetricQuotientN_{\indexI\indexJ}(\dpl^2_{\indexJ}) + \mathrm{T.E.},
\end{equation}
where T.E. is the truncation error which is approximately proportional to $\dpl^2$~\cite{PRESS2007_Recipe}. This means that the symmetric difference quotient function $\symmetricQuotientN_{\indexI\indexJ}$ is approximately linear in $\dpl^2$, thus, we represent it as a function of $\dpl^2$. The rounding error of numerical differentiation is omitted in the equation but it starts dominating over the truncation error below a certain value of $\dpl$. The truncation and rounding errors of the numerical integration which accumulate in $\localPars^{\indexJ}_{\final+\leftBraIdx\indexI\rightBraIdx}$  and $\localPars^{\indexJ}_{\final-\leftBraIdx\indexI\rightBraIdx}$ also exist even though they are not visible in the equation.

\cite{PRESS2007_Recipe, RIDDERS1982_DIFF} introduce a method to approximate the derivatives with a minimal truncation error of the numerical differentiation, i.e. $\symmetricQuotientN_{\indexI\indexJ}(0)$,  by extrapolating the values of $\symmetricQuotientN_{\indexI\indexJ}(\dpl^2)$ from iterations. The size of $\dpl$ decreases by a factor of $\riddersRate$ with each iteration but its initial value is set large enough to neglect the rounding error during the iterations. A polynomial extrapolation called Neville's algorithm~\cite{PRESS2007_Recipe, NEVILLE1934_Extrapolation} is performed using the results of the previous iteration and a lower order of the current iteration. The extrapolation is repeated recursively to reach a polynomial order equal to the iteration number, thus, the results of the numerical differentiation and polynomial extrapolations can be represented with an upper triangle matrix $\riddersMatrix^{\indexI\indexJ}$, illustrated as follows:

\begin{equation}
\renewcommand{\arraystretch}{1.5}
\begin{array}{|cc|cccccc|}
    \hline
    &  & \multicolumn{6}{c|}{\textrm{Iterations}} \\ 
    &  & 0 & 1 & 2 & 3 & \cdots & \riddersIteration \\ 
    \hline
   \parbox[t]{2mm}{\multirow{7}{*}{\rotatebox[origin=c]{90}{Orders}}} &    0 & \riddersMatrixIJ_{\leftBraIdx0,0\rightBraIdx} & \riddersMatrixIJ_{\leftBraIdx0,1\rightBraIdx} & \riddersMatrixIJ_{\leftBraIdx0,2\rightBraIdx} & \riddersMatrixIJ_{\leftBraIdx0,3\rightBraIdx} & \cdots & \riddersMatrixIJ_{\leftBraIdx0,\riddersIteration\rightBraIdx} 
   \\ 
   & 1 &  & \riddersMatrixIJ_{\leftBraIdx1,1\rightBraIdx} & \riddersMatrixIJ_{\leftBraIdx1,2\rightBraIdx} & \riddersMatrixIJ_{\leftBraIdx1,3\rightBraIdx} & & \multirow{4}{*}{\vdots}  \\ 
   & 2 & & & \riddersMatrixIJ_{\leftBraIdx2,2\rightBraIdx} & \riddersMatrixIJ_{\leftBraIdx2,3\rightBraIdx} & & \\
   & 3 &  & & & \riddersMatrixIJ_{\leftBraIdx3,3\rightBraIdx} &  & \\ 
   & \vdots &  & & & & \ddots & \\
   & \riddersIteration &  & & & & & \riddersMatrixIJ_{\leftBraIdx\riddersIteration,\riddersIteration\rightBraIdx}  
   \\ 
   \hline 
   \multicolumn{7}{c}{} \\ [-2.6ex]
\end{array}
\renewcommand{\arraystretch}{\defaultstretch} 
\end{equation}
The numerical differentiation at the first row and the polynomial extrapolations are conducted with the following equation:
\begin{equation}\label{eq:U_estimation}
    \riddersMatrixIJ_{\leftBraIdx\indexK,\indexL\rightBraIdx} = 
    \begin{dcases}
    \frac{\localPars^{\indexJ}_{\final+\leftBraIdx\indexI\rightBraIdx} - \localPars^{\indexJ}_{\final-\leftBraIdx\indexI\rightBraIdx}}{2\riddersRate^{-\indexL}\dpl_\indexJ} = \symmetricQuotientN_{\indexI\indexJ}(\dpl^2_{\indexJ}/\riddersRate^{2\indexL}) & \textrm{if } k = 0 \\
    \frac{\riddersMatrixIJ_{\leftBraIdx\indexK-1,\indexL\rightBraIdx}\riddersRate^{2\indexK} - \riddersMatrixIJ_{\leftBraIdx\indexK-1,\indexL-1\rightBraIdx}}{\riddersRate^{2\indexK} - 1} & \textrm{if } \indexK > 0
    \end{dcases}
\end{equation}
The $\numDiffMatrix_{\leftBraIdx\indexI, \indexJ\rightBraIdx}$, which is the optimal estimation of $\jacobianFull_{\leftBraIdx\indexI,\indexJ\rightBraIdx}$, is considered to be $\riddersMatrixIJ_{\leftBraIdx\indexK,\indexL\rightBraIdx}$ with $\indexK>0$ whose difference from $\max(\riddersMatrixIJ_{\leftBraIdx\indexK-1,\indexL\rightBraIdx},\riddersMatrixIJ_{\leftBraIdx\indexK-1,\indexL-1\rightBraIdx})$ is the smallest. The iteration completes at the $\riddersCompleteIndex$-th iteration when $\abs{\riddersMatrixIJ_{\leftBraIdx\riddersCompleteIndex-1,\riddersCompleteIndex-1\rightBraIdx} - \riddersMatrixIJ_{\leftBraIdx\riddersCompleteIndex,\riddersCompleteIndex\rightBraIdx}}$ is larger than the current smallest difference multiplied by a safety factor. In our test, $\dpl_\indexJ$ is set to \SI{1}{\mm} for $\ulocal$ and $\vlocal$, \SI{20}{mrad} for $\azimu$, \SI{1}{mrad} for $\polar$, and \SI{e-3}{\clight/\giga\eV} for $\qop$. The $\riddersRate$ is set to 1.1 and the safety factor to 5, which makes most of numerical differentiation complete in several dozen iterations.

\begin{figure*}[ht]
    \begin{center}
        \includegraphics[width=\jaocobiCompareFigureSize\linewidth]{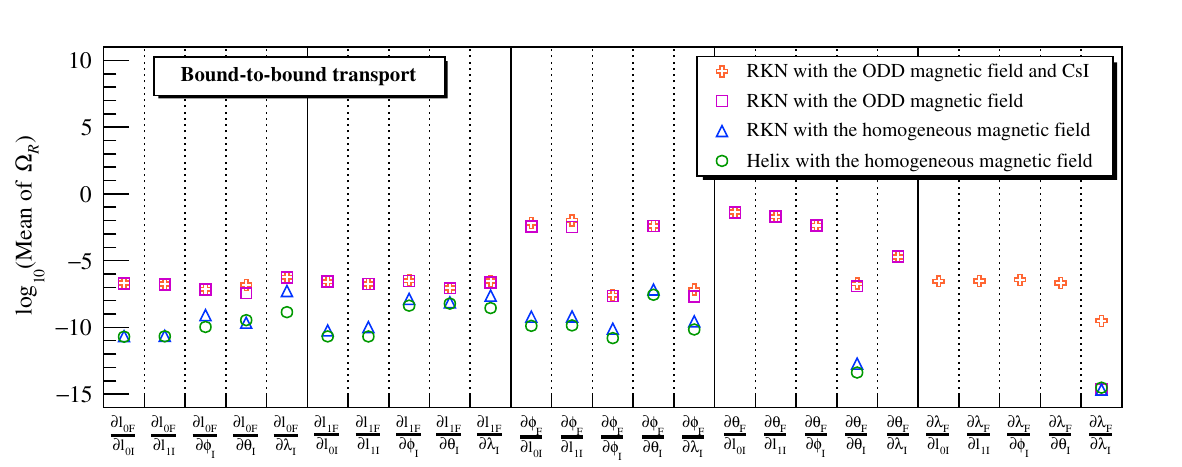}
        \includegraphics[width=\jaocobiCompareFigureSize\linewidth]{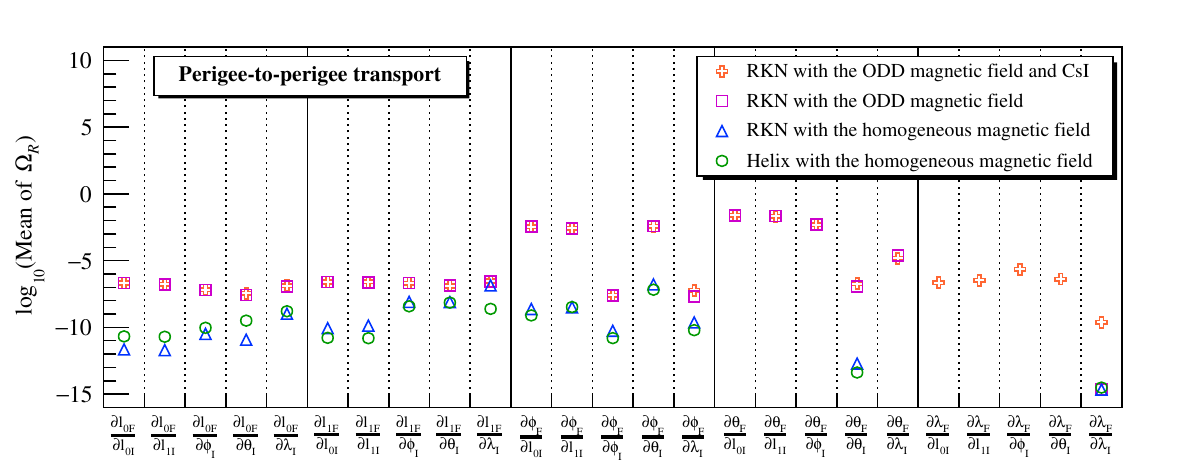}        
    \end{center}        
    \caption{Mean value of $\absRelResidual$ for each of the $\jacobianFull$ elements from (top) bound-to-bound transport and (bottom) perigee-to-perigee transport with different setups of magnetic fields and materials. The $\globalError$ of the reference tracks and the shifted tracks are set to $\toleranceMin$~mm and $\toleranceShitedTracks$~mm, respectively.}
    \label{fig:residuals_all_setups}
\end{figure*}

The metric to estimate the level of agreement between a Jacobian matrix element and the numerical differentiation results, is defined as the residual normalized by the numerical differentiation ($\absRelResidual$) \cite{LUND2009_CovarianceTransport}:
\begin{equation}\label{eq:absRelResidual}
    \absRelResidual^{\indexI \indexJ} = \left| \frac{\jacobianFull_{\leftBraIdx\indexI, \indexJ\rightBraIdx}-\numDiffMatrix_{\leftBraIdx\indexI, \indexJ\rightBraIdx}}{\numDiffMatrix_{\leftBraIdx\indexI, \indexJ\rightBraIdx}} \right|.  
\end{equation}
The mean values of $\absRelResidual^{\indexI \indexJ}$ for all elements of $\jacobianFull$ are measured with respect to $\globalError$, as shown in \cref{fig:residuals_full_setup}. $\jacobianFull$ is evaluated by propagating reference tracks for each value of $\globalError$ ranging from $\toleranceMin$~\si{\mm} to $\toleranceMax$~\si{\mm}. The $\globalError$ smaller than or equal to $\toleranceLimit$~\si{\mm} is not studied because a large fraction of tracks start failing in finding the final frame due to the precision limit of floating point operation. For the evaluation of $\numDiffMatrix$, even though it would be more precise to use the smallest $\globalError$ in the propagation of shifted tracks, $\toleranceShitedTracks$~\si{\mm} is used to reduce the computational cost. We observe that $\jacobianFull$ and $\numDiffMatrix$ have better agreement with smaller values of $\globalError$ due to smaller truncation errors from the numerical integration. For large values of $\globalError$, a few derivatives are found to be numerically unstable with residuals increasing up to the scale of $\mathcal{O}(1)$. The discrepancy becomes large for the Jacobian elements with infinitesimally small values which are sensitive to the numerical errors caused by magnetic field interpolation and numerical integration methods. For example, \cref{fig:residual histogram} shows that $\pdiff{\polarFinal}{\ulocalInitial}$, one of the most unstable derivatives, has a size smaller than $10^{-5}$~\SI{}{\per\mm} and $\pdiff{\qopFinal}{\qopInitial}$, the most stable derivative, is close to unity. The discrepancy of the extremely small Jacobian elements is tolerable because their impact on the final covariance matrix is negligible.

The residuals are also measured with two additional configurations to investigate the robustness of the estimation of Jacobian matrices against computational complexities due to gradients in the magnetic fields and material interactions. The first of these configurations comprises the ODD magnetic field without material, and the second comprises a homogeneous magnetic field with a strength of \SI{1.996}{\tesla} in the $z$-axis without material. For the configuration with the homogeneous magnetic field, the residuals are also evaluated using the analytical helix model~\cite{WITTEK1980_Helix, WITTEK1981_Helix, STRANDLIE2006_Helix} (see \ref{appendix:helix}) as well as the RKN method. The helical path length from $\globalParsInitial$, which satisfies \cref{eq:bound_implicit} or \cref{eq:perigee_implicit} at the final frame, is found using the Newton-Raphson method~\cite{PRESS2007_Recipe}, an iterative root-finding algorithm. The helical path length is updated by subtracting the ratio of $\implicitF(\globalPars)$ to its derivative with respect to the current helical path length:
\begin{equation}\label{eq:newton_method}
    \pathlength_{\indexNewton+1} = \pathlength_{\indexNewton} - \frac{\implicitF(\globalPars_{\indexNewton})}{\diffquote{\implicitF}(\globalPars_{\indexNewton})},
\end{equation}
where $\pathlength_{\indexNewton}$ is the helical path length of the $\indexNewton$-th iteration, and $\globalPars_{\indexNewton}$ are the global track parameters at $\pathlength_{\indexNewton}$. The initial path length, i.e. $\pathlength_0$, is evaluated using the straight, tangential ray from $\globalParsInitial$. The calculation of \cref{eq:newton_method} is repeated until the difference between $\pathlength_{\indexNewton+1}$ and $\pathlength_{\indexNewton}$ becomes less than \SI{e-9}{\mm}. The reference track and the corresponding shifted tracks are propagated in the same configuration with the same propagation method.

The mean values of $\absRelResidual$ for each configuration are shown in \cref{fig:residuals_all_setups} when propagating the $\nTracks$ reference tracks with $\tau$ of $\toleranceMin$~\si{\mm}. The residuals for the derivatives of $\polar$ with respect to the other track parameters are not obtained for the configuration with a homogeneous magnetic field because $\polar$ is invariant and the derivatives are zero, which makes \cref{eq:absRelResidual} unsolvable. The residuals for the derivatives of $\qop$ with respect to the other track parameters are also not obtained for the configurations without the material for the same reason. In the configurations without the material, $\pdiff{\qopFinal}{\qopInitial}$ is expected to be one from both $\jacobianFull$ and $\numDiffMatrix$, hence zero $\absRelResidual$. However, the numerical differentiation of \cref{eq:num_diff} makes slight offsets from one due to the rounding errors leading to non-zero $\absRelResidual$. For most elements of $\jacobianFull$, the precision decreases with the inhomogeneous magnetic field and the presence of the material does not noticeably impact the precision. As expected, the measurements with the fourth order RKN method and the analytical helix method show similar precision with the homogeneous magnetic field.

\subsection{Validation of covariance matrix transport}\label{subsec:cov_transport}

To validate \cref{eq:covariance_transport} statistically, a set of local track parameters at the initial frame smeared from $\localParsInitial$ are generated by sampling the displacements at random based on $\covarianceInitial$. The \textit{smeared tracks} are transported to the final frames to verify if the distributions of the local track parameters at the final frames follow $\covarianceFinal$ of the reference tracks. The validation is performed with the default configuration of the ODD magnetic field and a uniform CsI material distribution.

To generate the smeared track parameters, the covariance matrix is decomposed into a lower triangle matrix $\choleskydecompose$ and its transpose, namely:
\begin{equation}\label{eq:cholesky_decompose}
\covarianceInitial = \choleskydecompose \transpose{\choleskydecompose},
\end{equation}
where $\choleskydecompose$ can be calculated using the Cholesky decomposition. The smeared initial track parameters $\localParsInitialSmeared$ are generated by adding $\localParsInitial$ to the product of $\choleskydecompose$ and $\normalrandom$ containing elements randomly sampled from the normal distribution~\cite{PRESS2007_Recipe}:

\begin{equation}
    \localParsInitialSmeared = \localParsInitial + \choleskydecompose \normalrandom.
\end{equation}
The residual $\residual$ between the local track parameters at the final frame transported from $\localParsInitial$ and $\localParsInitialSmeared$ is calculated:

\begin{equation}
    \residual = \InitialToFinalMapping(\localParsInitialSmeared) - \InitialToFinalMapping(\localParsInitial) = \localParsFinalSmeared - \localParsFinal.
\end{equation}
Each component of $\residual$ is normalized by the standard deviations of $\localParsFinal$ to obtain a \textit{pull} ($\pullSymbol$)~\cite{LUND2009_CovarianceTransport}, also known as a standard score:

\begin{equation}
    \pull{\localPars_{\final\leftBraIdx\indexI\rightBraIdx}} = 
    \frac{\residual_{\leftBraIdx\indexI\rightBraIdx}}{\sqrt{\covariance_{\final\leftBraIdx\indexI,\indexI\rightBraIdx}}}.
\end{equation}

If the transport of the covariance matrix with the Jacobian matrix is a good approximation, the distributions of pulls should follow the normal distribution. To demonstrate this, we randomly produce a smeared track of $\localParsInitialSmeared$ once from a reference track. The reference and smeared tracks are propagated to the final frame to obtain the pulls. For every reference track, $\covarianceInitial$ is sampled at random before being decomposed by \cref{eq:cholesky_decompose}: the diagonal elements are set to the square of values randomly sampled from the Gaussian distributions with zero means and the standard deviations of \SI{50}{\um} for $\ulocalInitial$ and $\vlocalInitial$, \SI{1}{mrad} for $\azimuInitial$ and $\polarInitial$, and 10\% of $\qopInitial$ for $\qopInitial$, in accordance with the parameters used in \cite{LUND2009_CovarianceTransport}. This choice of the standard deviations is based on the specifications of the ATLAS experiment \cite{ATLAS_TDR}, which are conservative for muon momentum higher than around \SI[per-mode=symbol]{10}{\giga\eV\per\clight}. A correlation factor ($\correlation$) is sampled at random from the uniform distribution in the range of [-10\%,~10\%] and for every off-diagonal element of $\covarianceInitial$:

\begin{equation}
    \covariance_{\initial\leftBraIdx\indexI,\indexJ\rightBraIdx} = \covariance_{\initial\leftBraIdx\indexJ,\indexI\rightBraIdx} = \correlation^{\indexI\indexJ} \sqrt{\covariance_{\initial\leftBraIdx\indexI,\indexI\rightBraIdx} \covariance_{\initial\leftBraIdx\indexJ,\indexJ\rightBraIdx}}, \quad (\indexI \neq \indexJ).
\end{equation}

\begin{figure*}[ht]
    \begin{center}
    \includegraphics[width=\pullpvalFigureSize\linewidth]{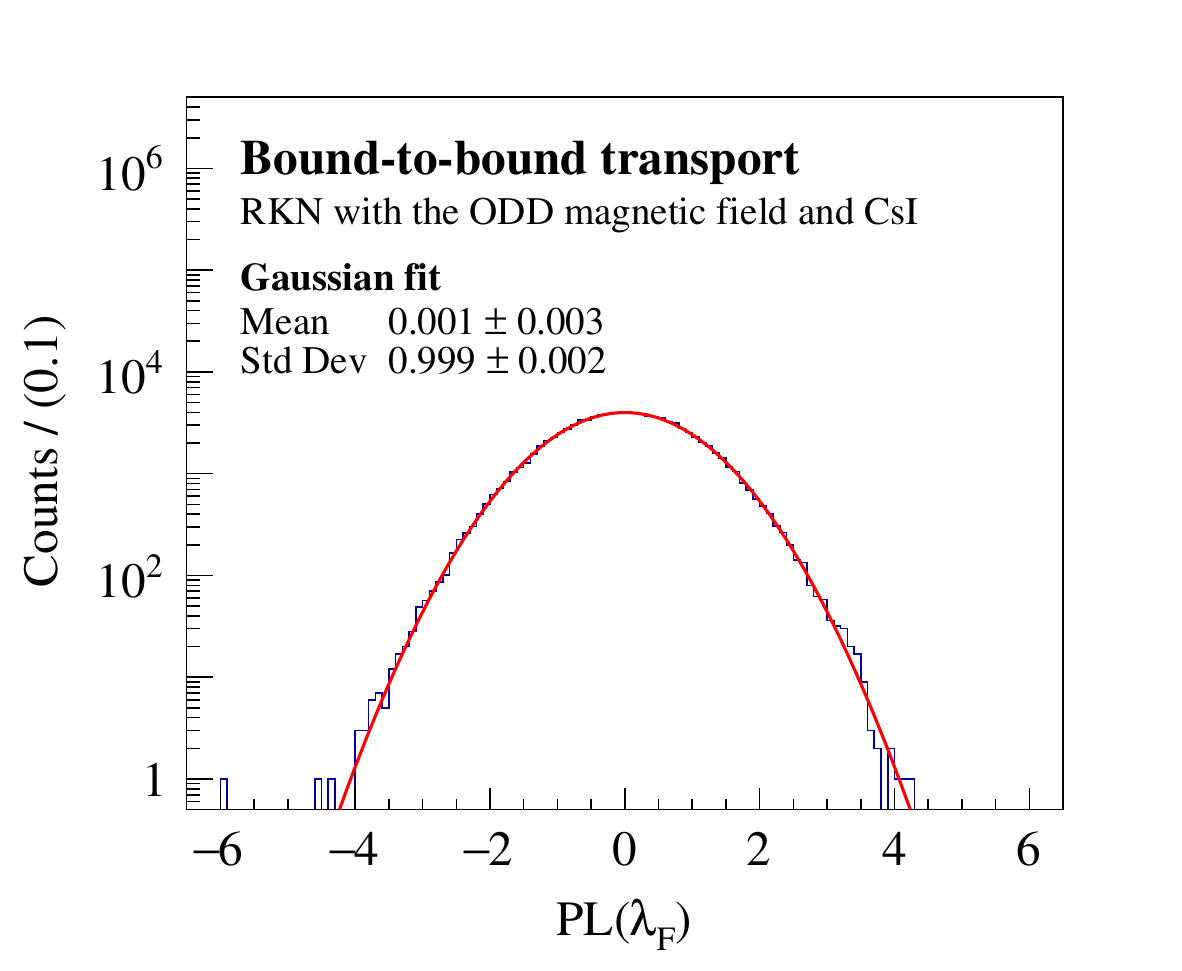}
    \includegraphics[width=\pullpvalFigureSize\linewidth]{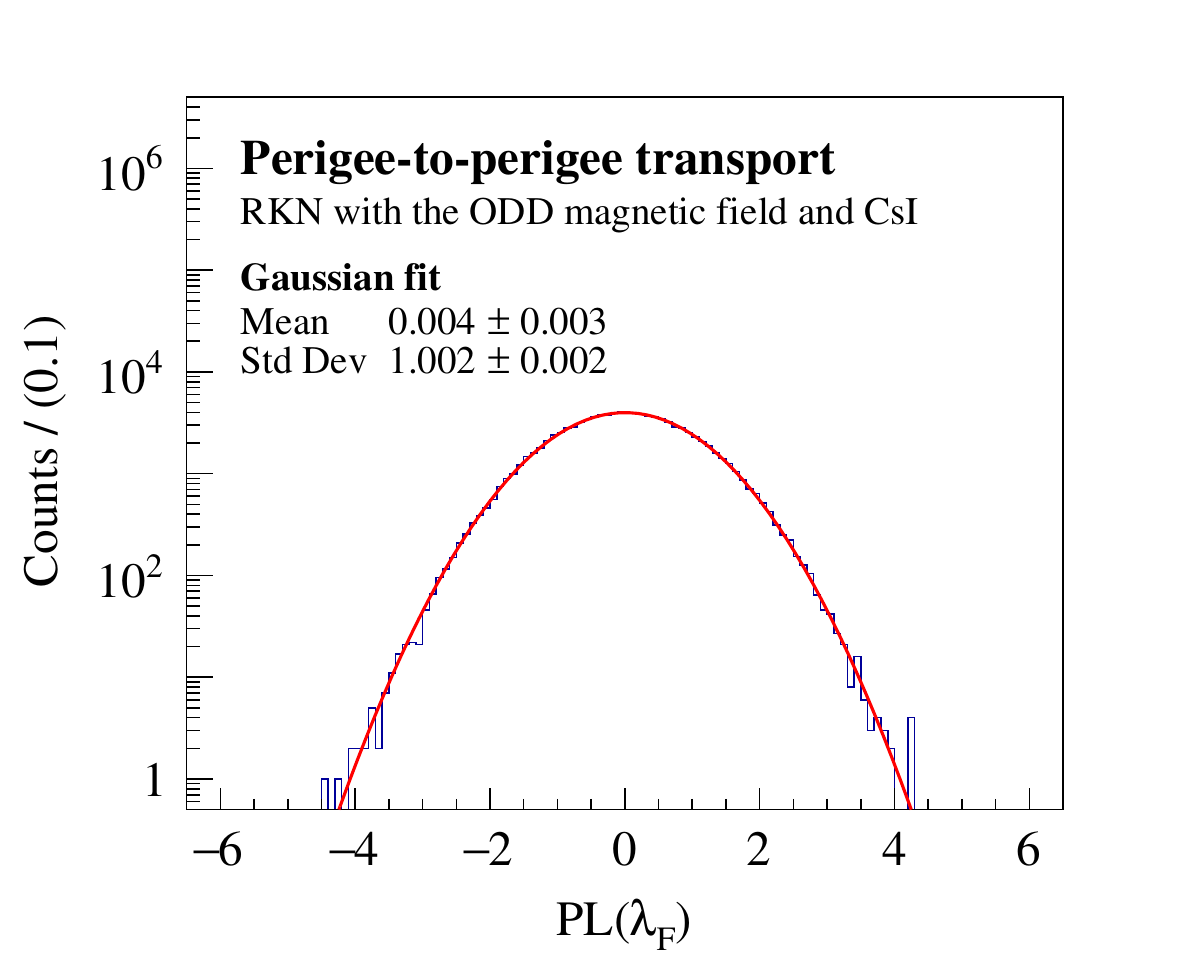}
    \caption{Pull distributions of $\qopFinal$ of (left) bound-to-bound transport and (right) perigee-to-perigee transport with the ODD magnetic field and CsI, fitted by a Gaussian function drawn in red line. The $\globalError$ of the reference tracks and the smeared tracks are set to $\toleranceCovTransport$~\si{\mm} and $\toleranceShitedTracks$~\si{\mm}, respectively.}
    \label{fig:pull_value_qop}
    \end{center}        
\end{figure*}

Once the $\nTracks$ reference and smeared tracks have been propagated, the pull distributions are fitted with Gaussian functions and their fitted parameters are listed in \cref{tab:pull_values}. As an example, the pull distributions of $\qopFinal$ with the Gaussian fit are also shown in \cref{fig:pull_value_qop}. For the propagation of the reference tracks including covariance matrix transport, the $\globalError$ of $\toleranceCovTransport$~\si{\mm} is used with feasible computational cost. For the smeared tracks, $\globalError$ is set to $\toleranceShitedTracks$~\si{\mm} as for the shifted tracks of numerical differentiation in \cref{subsec:numerical_diff}. The mean values and the standard deviations of the fitted Gaussian distributions are close to zero and one, respectively, indicating that the pulls follow the normal distribution. The pull values of $\azimuFinal$ have more outliers than expected from the normal distribution due to tracks almost parallel to the z-axis where the calculation of $\pdiff{\azimuFinal}{\dirFinal}$ of \cref{eq:n_w_derivatives} becomes numerically unstable due to the factor of $1/\sin{\polarFinal}$. 

\begin{figure*}[ht]
    \begin{center}
    \includegraphics[width=\pullpvalFigureSize\linewidth]{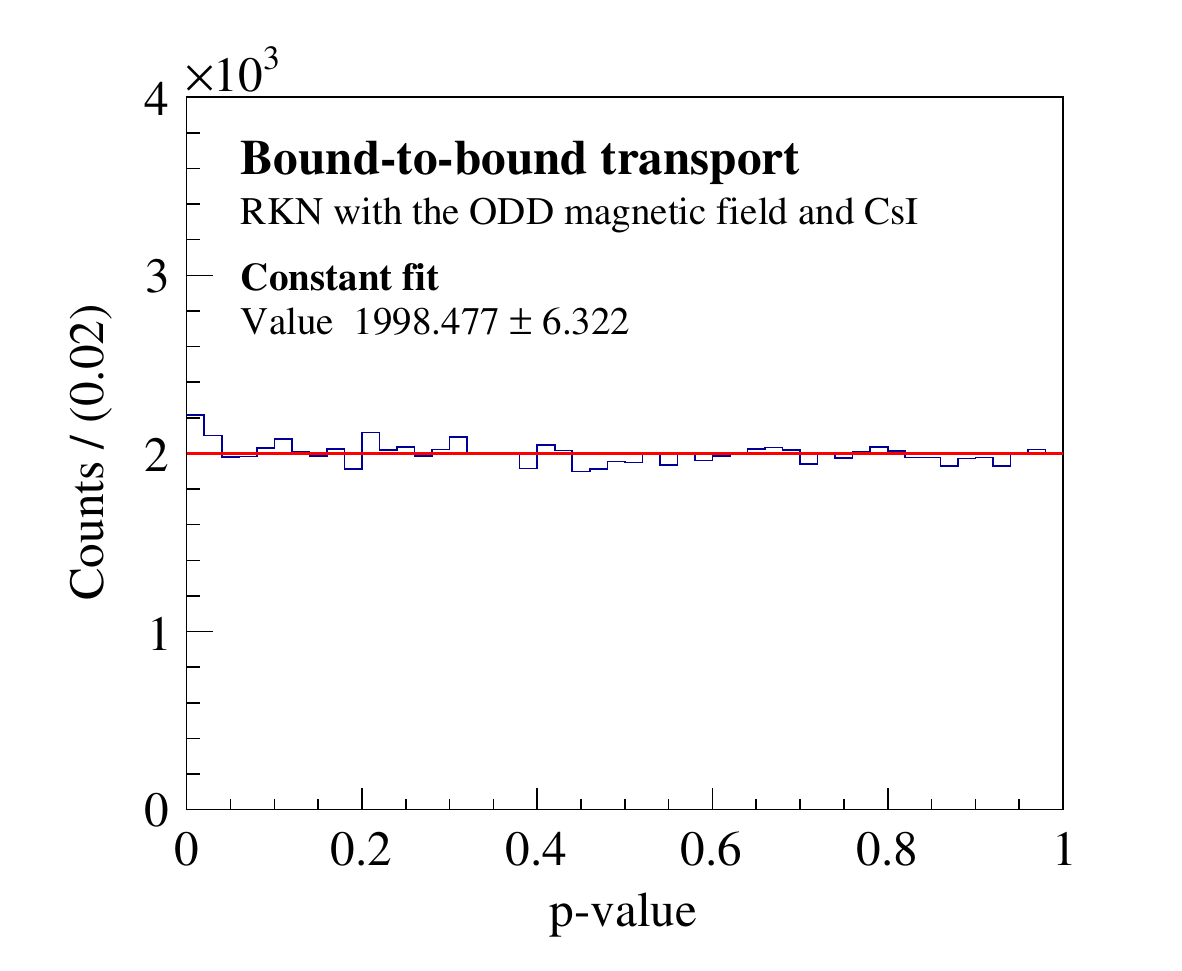}
    \includegraphics[width=\pullpvalFigureSize\linewidth]{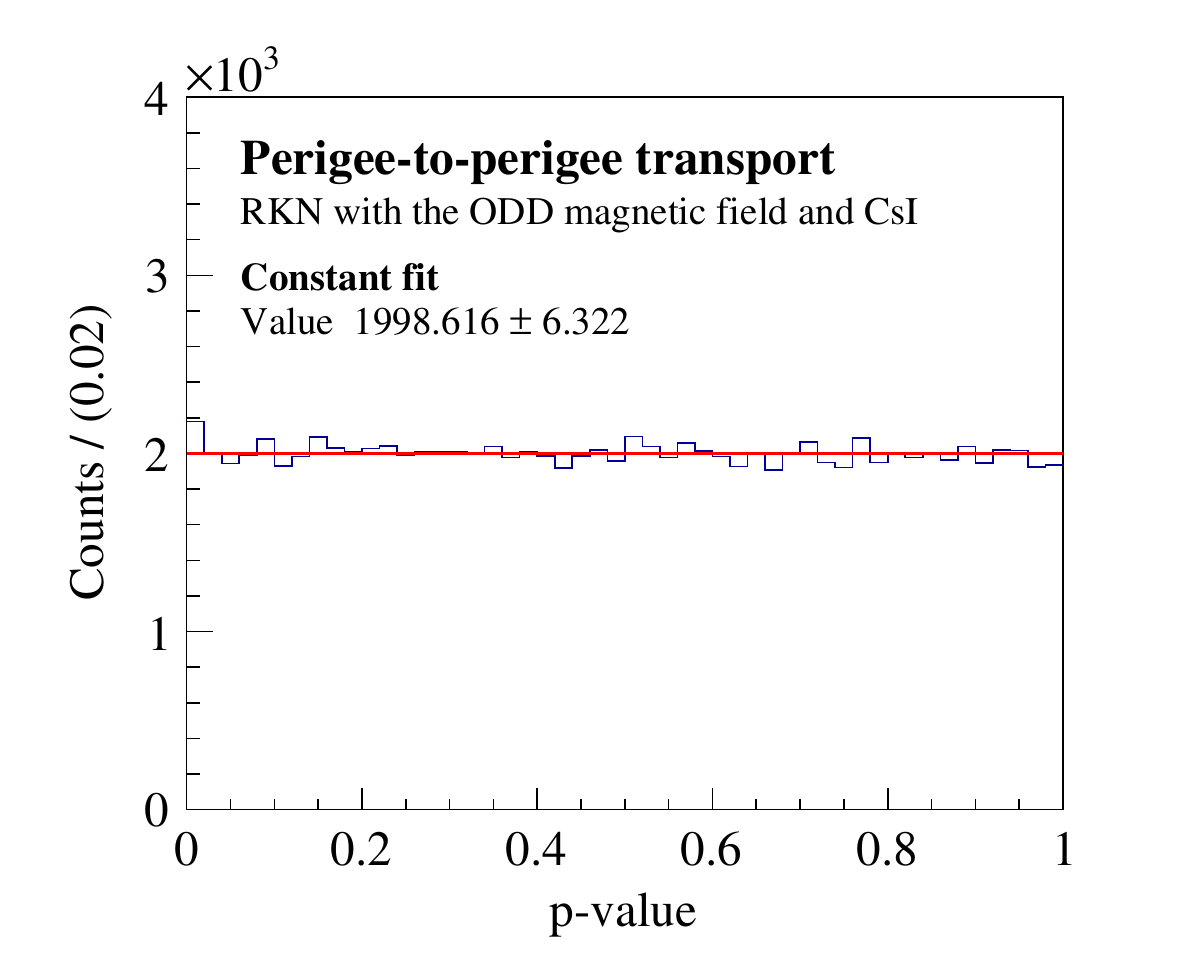}
    \caption{\pvalue-value distributions of (left) bound-to-bound transport and (right) perigee-to-perigee transport with the ODD magnetic field and CsI, fitted by a constant function drawn in red line. The $\globalError$ of the reference tracks and the smeared tracks are set to $\toleranceCovTransport$~\si{\mm} and $\toleranceShitedTracks$~\si{\mm}, respectively.}
    \label{fig:pvalue}
    \end{center}        
\end{figure*}

As a complementary method to qualify the covariance matrix transport with the Jacobian matrix, the randomness of \pvalue-value \cite{MURDOCH2008_PVALUE}, the probability that a sample at least as extreme as the current sample is obtained under the null hypothesis, is studied. The extremeness of the sample can be represented with the chi-square of $\residual$ which is given by:
\begin{equation}
    \chi^2 = \transpose{\residual} \inverse{\covarianceFinal} \residual. 
\end{equation}
The \pvalue-value is the upper tail of the chi-squared distribution with the number of degrees of freedom of local track parameters, i.e., five. It is equivalent to the unit subtracted by the value of its cumulative distribution function (CDF) at $\chi^2$ as follows:
\begin{equation}
 \textrm{\pvalue-value} = 1 - \operatorname{CDF}(\chi^2).
\end{equation}
The \pvalue-value is obtained for every reference track and its smeared track used for the pull distribution tests. The randomness of \pvalue-values can be confirmed from their distributions which almost follow the uniform distributions, as shown in \cref{fig:pvalue}. The small peaks at zero \pvalue-value are due to the linear approximation of the covariance matrix transport and the numerical instability in the track propagation.

\renewcommand{\arraystretch}{\defaultstretch}
\begin{table}[ht]
\small
\caption{Gaussian fitting results of pull distributions of tracks propagated for (left) bound-to-bound transport and (right) perigee-to-perigee transport with the ODD magnetic field and CsI. $\hat{\chi}^2$ is defined as a chi-square of the Gaussian fitting normalized by the number of degrees of freedom of the fitted data. $N_4$ means the number of the smeared tracks whose pull is larger than 4 or smaller than -4, and $\langle N_{4}\rangle$ is the expected number of such tracks under the assumption of the normal distribution, which is around $\nFourSigma$ for $\nTracks$ tracks. The $\globalError$ of the reference tracks and the smeared tracks are set to $\toleranceCovTransport$~\si{\mm} and $\toleranceShitedTracks$~\si{\mm}, respectively.}
\label{tab:pull_values}
\begin{center}
\begin{tabular}{ c r r c c  }
\toprule
Value & \multicolumn{1}{>{\centering\arraybackslash}p{2.3cm}}{Mean} & \multicolumn{1}{c}{Std Dev} & \multicolumn{1}{c}{$\hat{\chi}^2$} & \multicolumn{1}{c}{$N_{4}/\langle N_{4}\rangle$} \\ \midrule
\multicolumn{5}{l}{Bound-to-bound transport} \\  [0.2em]
$\pull{\ulocalFinal}$ & $0.001\pm0.003$ & $1.000\pm0.002$ & 1.36 & 0.79  \\ \midrule[0pt]
$\pull{\vlocalFinal}$ & $0.006\pm0.003$ & $1.004\pm0.002$ & 1.54 & 0.95  \\ \midrule[0pt]
$\pull{\azimuFinal}$  & $0.001\pm0.003$ & $0.999\pm0.002$ & 1.11 & 5.53  \\ \midrule[0pt]
$\pull{\polarFinal}$  & $0.001\pm0.003$ & $0.999\pm0.002$ & 0.70 & 1.26  \\ \midrule[0pt]
$\pull{\qopFinal}$    & $0.001\pm0.003$ & $0.999\pm0.002$ & 0.97 & 0.95  \\ 
\midrule
\multicolumn{5}{l}{Perigee-to-perigee transport} \\ [0.2em]
$\pull{\ulocalFinal}$ & $0.002\pm0.003$  & $1.002\pm0.002$ & 1.18 & 1.11 \\ \midrule[0pt]
$\pull{\vlocalFinal}$ & $0.000\pm0.003$  & $1.000\pm0.002$ & 1.16 & 1.26 \\ \midrule[0pt]
$\pull{\azimuFinal}$  & $-0.008\pm0.003$ & $1.001\pm0.002$ & 0.82 & 4.58 \\ \midrule[0pt]
$\pull{\polarFinal}$  & $-0.007\pm0.003$ & $1.001\pm0.002$ & 1.04 & 1.42 \\ \midrule[0pt]
$\pull{\qopFinal}$    & $0.004\pm0.003$  & $1.002\pm0.002$ & 0.87 & 1.26 \\
\bottomrule
\end{tabular}
\end{center}
\end{table}

\section{Discussion}\label{sec:discussion}

For the muon stopping power in the tests of \cref{sec:validation}, we use the mean value of energy loss from the collision process whose distribution follows the Landau distribution~\cite{Landau1944_EnergyLoss, Vavilov1957_EnergyLoss}. The Landau distribution of energy deposit is highly skewed right, which means that the mode, i.e. most probable value, is lower than the mean value. The shift of the mean is caused by rare collisions with high energy transfer, which corresponds to the upper tail of the energy loss distribution. In experiments where the material distribution is not very dense, it is more realistic to use the mode because it would be much closer to the actually observable mean energy loss than the mean, as suggested by~\cite{WORKMAN2022_PDG}. Adapting the work presented here to the mode would be straightforward because its equation~\cite{BICHSEL1988_MostProbableLoss} and derivative are similar to those of the Bethe equation introduced in \ref{appendix:mean_eloss}. 

For material interactions, only the inelastic collisions are considered, but Coulomb multiple scattering and radiative energy loss from Bremsstrahlung can also be important for low momenta and electrons, respectively~\cite{WORKMAN2022_PDG}. The implementation of multiple scattering into Jacobian matrices is challenging due to its randomness, which results in statistical fluctuations. The impact can be approximated by calculating an additional covariance matrix, assuming that the probability density functions (PDF) of the affected track parameters can be represented as a single or mixture of multiple Gaussian distributions~\cite{HIGHLAND1975_MS,LYNCH1991_MS}. The additional covariance matrix from the multiple scattering can be obtained for every step of the numerical integration and added to the covariance matrix of track parameters\footnote{It is also possible to calculate a single additional covariance matrix~\cite{FRUHWIRTH2000_DataAnalysis} of multiple scattering for the local-to-local frame transport if the material distribution is homogeneous. This is computationally cheaper but less precise in case of large energy loss.}. The implementation of Bremsstrahlung is relatively easy because its stopping power is approximately $\frac{\energy}{\radiationLength}$~\cite{ROSSI1952_Book} for electrons, where $\radiationLength$ is the radiation length of the material. 

The tests of \cref{sec:validation} validate the Jacobian matrix derivation in the various configurations of magnetic fields and materials but only the inhomogeneous material distribution is not tackled. Even though the derivation of \cref{sec:derivation} can be applied to any configuration, it is not recommended to perform the error propagation over non-uniform materials composition where the stopping power over the space is not continuous hence not differentiable. However, in real experiments, the material distribution is typically not homogeneous. One of the most representative cases is pixel detectors where a particle propagating in the vacuum enters the readout sensor and propagate further inside the sensor, which might have complicated material composition, to reach the point where the measurement occurs. In general, the situation can be simplified by assuming that the geometry component has a zero thickness---the thin scatterer approximation~\cite{FRUHWIRTH2021_PatternRecognition, FRUHWIRTH2000_DataAnalysis}---where a covariance matrix corresponding to the energy loss and multiple scattering from the original thickness is calculated and added to $\covarianceFinal$ instead of transporting it with a Jacobian matrix. The additional covariance matrix from the multiple scattering is calculated with the Gaussian distributions as explained in the previous paragraph. The method of additional covariance matrix can also be applied to the energy loss because the PDF of energy loss from the collision process follows the Landau distribution, which can be approximated into a Gaussian distribution, and the PDF of Bremsstrahlung energy loss can also be approximated into a mixture of multiple Gaussian distributions~\cite{FRUHWIRTH2003_GSF} like the multiple scattering. In cases where the thin scatterer assumption no longer holds, for example, particles traversing big chunks of materials such as calorimeters, the error propagation can be performed between volume boundaries where each volume is defined for a material chunk. As long as the volume boundaries are smooth enough, it is possible to define a bound frame tangential to the boundary surface where our formalism of local-to-local frame transport can be applied out of the box.

Only the Cartesian coordinate for the local bound frame has been discussed but detectors can produce measurements with a different coordinate system, e.g., Inner tracker strip sensor of the Phase II upgrade of the ATLAS experiment ~\cite{HESSEY2013_Annulus} with a polar coordinate. The additional calculation is done by either transforming the measurement and its error matrix into the local Cartesian coordinate or transforming $\local$ to the measurement coordinate system and multiplying the  additional coordinate transform Jacobian matrix to $\jacobianGlobalToLocal$ and $\jacobianLocalToGlobal$.

As indicated by \cref{fig:mean_step_size} and \cref{fig:residuals_all_setups}, the smaller values of $\globalError$ are more accurate but require more computing time in track propagation due to smaller step sizes. Standardizing the value of $\globalError$ for all experiments with different setups is not easy, however, there is a standard for fitting resolutions of track parameters required by experiments for the final physics analysis. This can be a guideline to investigate the optimized value of $\globalError$ once a track fitting algorithm is implemented. The computation with the 32-bit floating point numbers can also be attempted to save the computing time if the precision degradation is not significant.

\section{Summary and conclusions}\label{sec:summary}

The propagation of track parameters uncertainties is needed for track fitting in high-energy physics experiments. As the motion of charged particles in magnetic fields is described by a non-linear differential equation, the uncertainty propagation requires the evaluation of Jacobian matrices. These Jacobian matrices are constructed from gradients of track parameters between two local coordinate frames, each associated with a measurement of the corresponding particle. The Jacobian matrices for the propagation of helical tracks---i.e.\ tracks in homogeneous magnetic fields---have been derived in previous work. However, the Jacobian matrices for track propagation using numerical integration in inhomogeneous magnetic fields and material have not been derived completely in any previous work.

We have derived the Jacobian matrices for track propagation between local coordinate frames which can be used for arbitrary local coordinate frames and a range of numerical integration methods. We illustrate how this can be used for the local coordinate frame types widely used in the current collider experiments, namely bound and perigee frames. The total Jacobian matrix is decomposed into sub-Jacobian matrices corresponding to the coordinate transform between global and local coordinate system and track propagation in the global coordinate system. Such decomposition allows for the detailed derivation for different local frames and numerical integration methods.

The derivation has been validated by simulating track propagation between two frames with an inhomogeneous magnetic field and uniform material using fourth order RKN numerical integration. The Jacobian matrices evaluated from reference tracks are compared to the results of numerical differentiation obtained by shifting the reference tracks using a small change in the track parameters. Good agreement for most elements of Jacobian matrices was found, with the exception of a few numerically unstable ones. The test was repeated with simpler configurations---i.e.\ with a homogeneous magnetic field or without material---which obtained even better agreement. The covariance matrices transported by the Jacobian matrices are also validated against the transported track parameters at the final frame using tracks with initial parameters randomly smeared based on the initial covariance matrix. The pull distributions, i.e.\ the standard scores, of transported track parameters are well-described by the normal distribution. The \pvalue-value distributions have a small peak at zero due to the non-linearity and numerical instability of track propagation but can be approximately fit by the uniform distribution to verify the randomness of \pvalue-values.

Full track fitting would require additional implementation depending on the experimental setup. For example, although we use the mean energy loss from the Bethe equation, the most probable energy loss might be more realistic than the mean energy loss, which requires further study. To extend momentum range and include additional types of particles, multiple scattering and radiative energy loss would also need to be included. The evaluation of the Jacobian matrix can be cumbersome with inhomogeneous materials where the material property over the space is not continuous. The geometry could be partitioned into volumes each consisting of a single material or using the thin scatterer assumption if applicable. Measurements in non-Cartesian coordinate systems may need additional coordinate transforms to be used these results which are based on the Cartesian coordinate system. Numerical details, such as the error tolerance of the numerical integration or whether to use 32-bit or 64-bit floating point numbers, need to be studied for each experiment by investigating the computing time and the impact on the final track fitting results. 

\section*{Acknowledgements}
We would like to thank Prof. Are Strandlie for valuable comments and encouraging us to publish the results. This work was supported by the National Science Foundation under Cooperative Agreements OAC-1836650, PHY-2323298 and PHY-2120747. 

\appendix

\section{Track propagation with a helix model}\label{appendix:helix}

This provides the equations for the helix propagation model in the global coordinate system and its Jacobian matrices~\cite{WITTEK1980_Helix, WITTEK1981_Helix, STRANDLIE2006_Helix} to calculate $\bigl(\pdiff{\globalParsFinal}{\globalParsInitial}\bigr)_\pathlength$ of \cref{eq:jac_g2g}. The helix model can be used as an approximation for experiments where the gradients of magnetic field is not large and the material interactions are negligible.

We start by defining the global track parameters at $\pathlength = 0$ as ${\transpose{\globalParsZero}=(\transpose{\posZero} \mquad \transpose{\dirZero} \mquad \qopHelix)}$. The global position $\posHelix$ along the helical trajectory in the constant magnetic field $\bfield$ is given as a function of $\pathlength$ ~\cite{WITTEK1980_Helix, WITTEK1981_Helix, STRANDLIE2006_Helix}:
\begin{align}\label{eq:helix_pos}
    \posHelix = \posZero & + \frac{\QtermHelix\pathlength - \sin{(\QtermHelix\pathlength)}}{\QtermHelix} (\bfieldNormalized \cdot \dirZero)\bfieldNormalized \nonumber \\
    & + \frac{\sin{(\QtermHelix\pathlength)}}{\QtermHelix} \dirZero + \frac{1 - \cos{(\QtermHelix\pathlength)}}{\QtermHelix} (\bfieldNormalized \cross \dirZero).
\end{align}
where $\QtermHelix = -|\bfield|\qopHelix$ and $\bfieldNormalized = \frac{\bfield}{|\bfield|}$. The direction $\dirHelix$ is simply the derivative of $\pos$ with respect to $\pathlength$:

\begin{align}\label{eq:helix_dir}
    \dirHelix & = [1 - \cos{(\QtermHelix\pathlength)}] (\bfieldNormalized \cdot \dirZero) \bfieldNormalized \nonumber \\ & \qquad + \cos{(\QtermHelix\pathlength)} \cdot \dirZero + \sin{(\QtermHelix\pathlength)} (\bfieldNormalized \times \dirZero).
\end{align}
The Jacobian matrix $\pdiff{\globalPars}{\globalParsZero}$ is represented with block matrices:
\begin{equation}\label{eq:dgdg0_helix}
\pdiff{\globalParsHelix}{\globalParsZero} =    
    \begin{pmatrix}[\matrixSpacingBig]
     \scalebox{\pdiffScaleFactor}{\pdiff{\posHelix}{\posZero}} & \scalebox{\pdiffScaleFactor}{\pdiff{\posHelix}{\dirZero}} & \scalebox{\pdiffScaleFactor}{\pdiff{\posHelix}{\qopHelix}} \\
     \zeroMatrix{3}{3} & \scalebox{\pdiffScaleFactor}{\pdiff{\dirHelix}{\dirZero}} & 
     \scalebox{\pdiffScaleFactor}{\pdiff{\dirHelix}{\qopHelix}}\\
     \zeroMatrix{1}{3} & \zeroMatrix{1}{3} & \identityMatrix{1}
    \end{pmatrix}.
\end{equation}

The following identities are useful in differentiating ${(\bfieldNormalized \cdot \dirZero)\bfieldNormalized}$ with respect to $\dirZero$:
\begingroup
\addtolength{\jot}{\equationSpacingA}
\begin{align}\label{eq:derivative_of_BBN}
    (\bfieldNormalized \cdot \dirZero) \bfieldNormalized & = \bfieldNormalized \transpose{\bfieldNormalized} \dirZero, \nonumber \\
    \pdiff{}{\dirZero}[(\bfieldNormalized \cdot \dirZero) \bfieldNormalized] & = \bfieldNormalized \transpose{\bfieldNormalized}.
\end{align}
\endgroup
To differentiate $(\bfieldNormalized \times \dirZero)$ with respect to $\dirZero$, the column-wise matrix cross product ($\columnwiseCrossProd$) is defined using a cross product of each column of matrix with a given vector. For example, a $3\times3$ matrix $\mathbf{X}$ with the columns of $\mathbf{x}_0, \mathbf{x}_1$ and $\mathbf{x}_2$, has a column-wise matrix cross product with a three-element vector $\mathbf{a}$ defined as follows:
\begin{align}\label{eq:columnwise_cross_product}
    \mathbf{X} \columnwiseCrossProd \mathbf{a} & =     
    \begin{pmatrix}
     \mathbf{x}_0 & \mathbf{x}_1 & \mathbf{x}_2  
    \end{pmatrix}    
    \columnwiseCrossProd \mathbf{a} \nonumber \\ 
    & = \begin{pmatrix}
    \mathbf{x}_0 \times \mathbf{a} & \mathbf{x}_1 \times \mathbf{a} & \mathbf{x}_2 \times \mathbf{a}
    \end{pmatrix}.
\end{align}
The column-wise matrix cross product can be applied to the derivative of the cross product between two three-element vectors with respect to another three-element vector:
\begin{equation}\label{eq:columnwise_cross_product_chainrule}
    \pdiff{(\mathbf{x}_0 \times \mathbf{x}_1)}{\mathbf{a}} = \pdiff{\mathbf{x}_0}{\mathbf{a}} \columnwiseCrossProd \mathbf{x}_1 - \pdiff{\mathbf{x}_1}{\mathbf{a}} \columnwiseCrossProd \mathbf{x}_0,
\end{equation}
which directly provides the following identity:
\begin{equation}\label{eq:derivative_of_BcrossN}
    \pdiff{}{\dirZero} (\bfieldNormalized \times \dirZero) = - \identityMatrix{3} \columnwiseCrossProd \bfieldNormalized.
\end{equation}

The non-zero block matrices of \cref{eq:dgdg0_helix} can be derived easily~\cite{WITTEK1981_Helix, STRANDLIE2006_Helix} using \cref{eq:derivative_of_BBN} and \cref{eq:derivative_of_BcrossN}:
\begingroup
\addtolength{\jot}{\equationSpacingB}
\begin{align}\label{eq:helix_jacobian}
\pdiff{\posHelix}{\posZero} & = \identityMatrix{3}, \\
\pdiff{\posHelix}{\dirZero} & = \frac{\sin{(\QtermHelix\pathlength)}}{\QtermHelix} \identityMatrix{3} \nonumber \\
& \quad + \frac{\QtermHelix\pathlength - \sin{(\QtermHelix\pathlength)}}{\QtermHelix} \bfieldNormalized \transpose{\bfieldNormalized} + \frac{\cos{(\QtermHelix\pathlength)-1}}{\QtermHelix}\identityMatrix{3} \columnwiseCrossProd \bfieldNormalized,  \\
\pdiff{\posHelix}{\qopHelix} & =\frac{1}{\qopHelix} (\pathlength \dirHelix + \posZero - \posHelix),  \\
\pdiff{\dirHelix}{\dirZero} & = \cos{(\QtermHelix\pathlength)}\identityMatrix{3} \nonumber \\
& \quad + [ 1 - \cos{(\QtermHelix\pathlength)}] \bfieldNormalized \transpose{\bfieldNormalized} - \sin{(\QtermHelix\pathlength)} \identityMatrix{3} \columnwiseCrossProd \bfieldNormalized,  \\
\pdiff{\dirHelix}{\qopHelix} & = \frac{\QtermHelix\pathlength}{\qopHelix}[\sin(\QtermHelix\pathlength)(\bfieldNormalized\transpose{\bfieldNormalized} - \identityMatrix{3})\dirZero + \cos(\QtermHelix\pathlength)\bfieldNormalized \times \dirZero] \label{eq:dndqop_helix}.
\end{align}
\endgroup
The $\pdiff{\dirHelix}{\qopHelix}$ of \cref{eq:dndqop_helix} corrects the equation of \cite{WITTEK1981_Helix, STRANDLIE2006_Helix} which only works when $\QtermHelix\pathlength \ll 1$. The $\pdiff{\globalParsHelix}{\globalParsZero}$ of \cref{eq:dgdg0_helix} can be used as $\bigl(\pdiff{\globalParsFinal}{\globalParsInitial}\bigr)_\pathlength$ of \cref{eq:jac_g2g} because the $\pathlength$ is an independent parameter. 

\section{Track propagation with the adaptive fourth order Runge-Kutta-Nystr{\"o}m method}\label{appendix:rkn4}

The application of the fourth order RKN method to the track propagation can be found in \cite{BUGGE1981_RK,LUND2009_CovarianceTransport, LUND2009_RKNPropagation}. This section briefly introduces how the fourth order RKN method advances global track parameters and provide some details on its numerical integration to calculate $\bigl(\pdiff{\globalParsFinal}{\globalParsInitial}\bigr)_\pathlength$ of \cref{eq:jac_g2g}.

\subsection{Advancing global track parameters per step}

The RKN method numerically integrates $\pos$ and $\dir$ with a given step size of $\stepsize$ accumulated in $\pathlength$ to solve the second order differential equation \cref{eq:EquationOfMotion}, namely $\scalarFunc$:
\begin{equation}
    \doubquote{\pos} = \scalarFunc(\pathlength, \pos,\diffquote{\pos}) = \scalarFunc(\pathlength, \pos, \dir),
\end{equation}
where the quotation mark denotes the partial derivative with respect to $\pathlength$, thus, $\diffquote{\pos}$ is equivalent to $\dir$. 

For every step of the fourth order RKN method, the global track parameters are calculated at four different stages to evaluate  \cref{eq:EquationOfMotion}, namely $\ddirds_{\indexRK}$ \cite{NYSTROM1925_RKN, ABRAMOWITZ1972_Handbook}:
\begingroup
\addtolength{\jot}{\equationSpacingA}
\begin{align}\label{eq:k_stages}
   \ddirds_1 & = \scalarFunc(s_1, \pos_1, \dir_1) =  \scalarFunc(s,\pos, \dir),  \nonumber \\
   \ddirds_2 & = \scalarFunc(s_2, \pos_2, \dir_2) = \scalarFunc(s + \frac{\stepsize}{2},\pos + \frac{\stepsize}{2} \dir + \frac{\stepsize^2}{8} \ddirds_1, \dir + \frac{\stepsize}{2} \ddirds_1), \nonumber \\
   \ddirds_3 & = \scalarFunc(s_3, \pos_3, \dir_3) = \scalarFunc(s + \frac{\stepsize}{2},\pos + \frac{\stepsize}{2} \dir + \frac{\stepsize^2}{8} \ddirds_1, \dir + \frac{\stepsize}{2} \ddirds_2), \nonumber \\
   \ddirds_3 & = \scalarFunc(s_4, \pos_4, \dir_4) = \scalarFunc(s + \stepsize,\pos + \stepsize \dir + \frac{\stepsize^2}{2} \ddirds_3, \dir + \stepsize \ddirds_3),
\end{align}
\endgroup
where $\pos$, $\dir$, and $\qop$ represent the global track parameters of the current step, and the parameters at the four stages of each step have the subscripts of 1, 2, 3, and 4. It should be noted that $\ddirds_{\indexRK}$ is evaluated as follows:
\begin{equation}
    \ddirds_{\indexRK} = \qop_{\indexRK} (\dir_{\indexRK} \cross \bfield_{\indexRK}).
\end{equation}

Before evaluating $\ddirds_{\indexRK}$, in the presence of materials, $\qop_{\indexRK}$ should be known by calculating $\qop$ at $\pathlength+\frac{\stepsize}{2}$ and $\pathlength+\stepsize$. We use the fourth order Runge-Kutta method~\cite{PRESS2007_Recipe} to evaluate them as done in~\cite{BUGGE1981_RK}:
\begingroup
\addtolength{\jot}{\equationSpacingA}
\begin{align}\label{eq:qop_stages}
    \qop_1 & = \qop, \nonumber \\
    \qop_2 & = \qop + \frac{\stepsize}{2}\dqopds_1, \nonumber \\
    \qop_3 & = \qop + \frac{\stepsize}{2}\dqopds_2, \nonumber \\
    \qop_4 & = \qop + \stepsize \dqopds_3,
\end{align}
\endgroup
where $\dqopds_{\indexRK}$ is given by:
\begin{equation}
    \dqopds_{\indexRK} = \pdiff{\qop_{\indexRK}}{\pathlength} = \frac{\qop^{3}_{\indexRK} \energy_{\indexRK}}{\charge^2} \left( -\pdiff{\energy_{\indexRK}}{\pathlength} \right).
\end{equation}

$\dir_{\indexRK}$ and $\pos_{\indexRK}$ are directly provided in \cref{eq:k_stages}, which can be recursively obtained by calculating $\ddirds_{\indexRK}$ of the previous stage. The positions, $\pos_2$ and $\pos_3$ are the same, as are $\bfield_2$ and $\bfield_3$.

The global track parameters for the next step $(\newstate{\pos}, \newstate{\dir}, \newstate{\qop})$ are estimated as follows~\cite{BUGGE1981_RK}:
\begingroup
\addtolength{\jot}{\equationSpacingA}
\begin{align}\label{eq:rkn_advancement}
    \newstate{\pos} & = \pos + \stepsize \dir + \frac{\stepsize^2}{6}(\ddirds_1 + \ddirds_2 + \ddirds_3), \nonumber \\
    \newstate{\dir} & = \dir + \frac{\stepsize}{6} (\ddirds_1 + 2\ddirds_2 + 2\ddirds_3 + \ddirds_4), \nonumber \\  
    \newstate{\qop} & = \qop + \frac{\stepsize}{6} (\dqopds_1 + 2\dqopds_2 + 2\dqopds_3 + \dqopds_4). 
\end{align}
\endgroup
In the validation tests of \cref{sec:validation}, $\newstate{\dir}$ is normalized for each step to ensure that it is always the unit vector.

The local error ($\localError$), the absolute difference between $\newstate{\pos}$ of the fourth and third order RKN method, is estimated as follows~\cite{LUND2009_RKNPropagation}:
\begin{equation}
    \localError = \left|\frac{\stepsize^2}{6}(\ddirds_1 - \ddirds_2 - \ddirds_3 + \ddirds_4 ) \right|.
\end{equation}
The next step size ($\newstate{\stepsize}$) is scaled with the ratio of the error tolerance and local error estimation~\cite{LUND2009_RKNPropagation}:
\begin{equation}
    \newstate{\stepsize} = \stepsize \left(\frac{\globalError}{\localError}\right)^{\frac{1}{4}},
\end{equation}
which is constrained between $\frac{\stepsize}{4}$ and $4\stepsize$ to prevent a drastic change in the step size.

\subsection{Calculation of $\left(\pdiff{\globalParsFinal}{\globalParsInitial}\right)_\pathlength$}
 We start by defining a step Jacobian matrix ($\jacobianStep$) for the propagation from $(\pos,\dir,\qop)$ to $(\newstate{\pos}, \newstate{\dir}, \newstate{\qop})$ corresponding to a single step of the RKN method:  
\begin{equation}
    \jacobianStep = \pdiff{(\newstate{\pos}, \newstate{\dir}, \newstate{\qop})}{(\pos, \dir, \qop)} =     \begin{pmatrix}[\matrixSpacingBig]
     \scalebox{\pdiffScaleFactor}{\pdiff{\newstate{\pos}}{\pos}} & \scalebox{\pdiffScaleFactor}{\pdiff{\newstate{\pos}}{\dir}} & 
     \scalebox{\pdiffScaleFactor}{\pdiff{\newstate{\pos}}{\qop}} \\
     \scalebox{\pdiffScaleFactor}{\pdiff{\newstate{\dir}}{\pos}} & \scalebox{\pdiffScaleFactor}{\pdiff{\newstate{\dir}}{\dir}} & 
     \scalebox{\pdiffScaleFactor}{\pdiff{\newstate{\dir}}{\qop}} \\
     \scalebox{\pdiffScaleFactor}{\pdiff{\newstate{\qop}}{\pos}} & \scalebox{\pdiffScaleFactor}{\pdiff{\newstate{\qop}}{\dir}} & \scalebox{\pdiffScaleFactor}{\pdiff{\newstate{\qop}}{\qop}} 
     \\[\lastRowMargin]
    \end{pmatrix}.
\end{equation}
If the track propagation makes $\RKstepsTotal$ steps to reach the final frame from the initial frame, $\bigl(\pdiff{\globalParsFinal}{\globalParsInitial}\bigr)_\pathlength$ can be integrated using the \textit{Bugge-Myrheim method}~\cite{BUGGE1981_RK, LUND2009_CovarianceTransport} as follows:
\begin{equation}
    \left(\pdiff{\globalParsFinal}{\globalParsInitial}\right)_\pathlength = \jacobianStep_{\RKstepsTotal} \jacobianStep_{\RKstepsTotal-1} \cdots \jacobianStep_{1},
\end{equation}
where $\jacobianStep_\indexI$ is the step Jacobian matrix of the $\indexI$-th step. 

Each block matrices of $\jacobianStep$ can be expanded using \cref{eq:rkn_advancement}:
\begingroup
\addtolength{\jot}{\equationSpacingB}
\begin{align}\label{eq:D_blocks}
    \pdiff{\newstate{\pos}}{\pos} & = \identityMatrix{3} + \frac{\stepsize^2}{6}\left(\pdiff{\ddirds_1}{\pos} + \pdiff{\ddirds_2}{\pos} + \pdiff{\ddirds_3}{\pos}\right), \nonumber \\
    \pdiff{\newstate{\pos}}{\dir} & = h \identityMatrix{3} + \frac{\stepsize^2}{6}\left(\pdiff{\ddirds_1}{\dir} + \pdiff{\ddirds_2}{\dir} + \pdiff{\ddirds_3}{\dir}\right), \nonumber \\
    \pdiff{\newstate{\pos}}{\qop} & = \frac{\stepsize^2}{6}\left(\pdiff{\ddirds_1}{\qop} + \pdiff{\ddirds_2}{\qop} + \pdiff{\ddirds_3}{\qop}\right), \nonumber \\
    \pdiff{\newstate{\dir}}{\pos} & = \frac{\stepsize}{6}\left(\pdiff{\ddirds_1}{\pos}+ 2\pdiff{\ddirds_2}{\pos} + 2\pdiff{\ddirds_3}{\pos} + \pdiff{\ddirds_4}{\pos}\right), \nonumber \\
    \pdiff{\newstate{\dir}}{\dir} & = \identityMatrix{3} + \frac{\stepsize}{6}\left(\pdiff{\ddirds_1}{\dir}+ 2\pdiff{\ddirds_2}{\dir} + 2\pdiff{\ddirds_3}{\dir} + \pdiff{\ddirds_4}{\dir}\right), \nonumber \\
    \pdiff{\newstate{\dir}}{\qop} & = \frac{\stepsize}{6}\left(\pdiff{\ddirds_1}{\qop}+ 2\pdiff{\ddirds_2}{\qop} + 2\pdiff{\ddirds_3}{\qop} + \pdiff{\ddirds_4}{\qop}\right), \nonumber \\
    \pdiff{\newstate{\qop}}{\pos} & = \frac{\stepsize}{6}\left(\pdiff{\dqopds_1}{\pos}+ 2\pdiff{\dqopds_2}{\pos} + 2\pdiff{\dqopds_3}{\pos} + \pdiff{\dqopds_4}{\pos}\right), \nonumber \\
    \pdiff{\newstate{\qop}}{\dir} & = \frac{\stepsize}{6}\left(\pdiff{\dqopds_1}{\dir}+ 2\pdiff{\dqopds_2}{\dir} + 2\pdiff{\dqopds_3}{\dir} + \pdiff{\dqopds_4}{\dir}\right),  \nonumber \\ 
    \pdiff{\newstate{\qop}}{\qop} & = 1 + \frac{\stepsize}{6}\left(\pdiff{\dqopds_1}{\qop}+ 2\pdiff{\dqopds_2}{\qop} + 2\pdiff{\dqopds_3}{\qop} + \pdiff{\dqopds_4}{\qop}\right).
\end{align}
\endgroup

$\pdiff{\ddirds_{\indexRK}}{\pos}$ can be calculated using the column-wise matrix cross product defined in \cref{eq:columnwise_cross_product}:

\begin{equation}\label{eq:dkndr}
\pdiff{\ddirds_{\indexRK}}{\pos} = \qop_{\indexRK} \left[\pdiff{\dir_{\indexRK}}{\pos} \columnwiseCrossProd \bfield_{\indexRK} - \left( \pdiff{\bfield_{\indexRK}}{\pos_{\indexRK}}\pdiff{\pos_{\indexRK}}{\pos} \right) \columnwiseCrossProd \dir_{\indexRK}  \right],
\end{equation}
where a chain rule is applied to $\pdiff{\bfield_{\indexRK}}{\pos}$ given that inhomogeneous magnetic fields are a function of the positions, and the magnetic field gradients at $\pos_{\indexRK}$ can be obtained using the numerical differentiation. The term with $\pdiff{\qop_{\indexRK}}{\pos}$ is neglected here, which is zero in the case of the homogeneous material configuration. This term is non-zero in the case a track propagates through inhomogeneous materials between two frames but such a configuration is rarely used in the experimental software. Typically, the covariance matrices are transported within the same material by dividing geometry volumes appropriately. 

\cref{eq:dkndr} can be expanded as follows using the expressions of $\pos_{\indexRK}$ and $\dir_{\indexRK}$ of \cref{eq:k_stages}:
\begingroup
\addtolength{\jot}{\equationSpacingB}
\begin{align}
    \pdiff{\ddirds_1}{\pos} & = -\qop_1 \pdiff{\bfield_1}{\pos_1} \columnwiseCrossProd \dir_1, \nonumber \\
    \pdiff{\ddirds_2}{\pos} & = \qop_2 \left\{ \frac{\stepsize}{2} \pdiff{\ddirds_1}{\pos} \columnwiseCrossProd \bfield_2 - \left[ \pdiff{\bfield_2}{\pos_2} \left( \identityMatrix{3} + \frac{\stepsize^2}{8} \pdiff{\ddirds_1}{\pos} \right) \right] \columnwiseCrossProd \dir_2 \right\}, \nonumber  \\
    \pdiff{\ddirds_3}{\pos} & = \qop_3 \left\{ \frac{\stepsize}{2} \pdiff{\ddirds_2}{\pos} \columnwiseCrossProd \bfield_3 - \left[ \pdiff{\bfield_3}{\pos_3} \left( \identityMatrix{3} + \frac{\stepsize^2}{8} \pdiff{\ddirds_1}{\pos} \right) \right] \columnwiseCrossProd \dir_3 \right\}, \nonumber \\
    \pdiff{\ddirds_4}{\pos} & = \qop_4 \left\{ \stepsize \pdiff{\ddirds_3}{\pos} \columnwiseCrossProd \bfield_4 - \left[ \pdiff{\bfield_4}{\pos_4} \left( \identityMatrix{3} + \frac{\stepsize^2}{2} \pdiff{\ddirds_3}{\pos} \right) \right] \columnwiseCrossProd \dir_4 \right\},
\end{align}
\endgroup
where $\pdiff{\ddirds_{\indexRK}}{\pos}$ can be calculated recursively. $\pdiff{\ddirds_{\indexRK}}{\dir}$ is calculated in the similar way:
\begin{equation}
\pdiff{\ddirds_{\indexRK}}{\dir} = \qop_{\indexRK} \left[\pdiff{\dir_{\indexRK}}{\dir} \columnwiseCrossProd \bfield_{\indexRK} -  \left( \pdiff{\bfield_{\indexRK}}{\pos_{\indexRK}}\pdiff{\pos_{\indexRK}}{\dir} \right) \columnwiseCrossProd \dir_{\indexRK}  \right],
\end{equation}
which is expanded for the recursive calculation:
\begingroup
\addtolength{\jot}{\equationSpacingB}
\begin{align}\label{eq:dknddir}
    \pdiff{\ddirds_1}{\dir} & = \qop_1 \identityMatrix{3} \columnwiseCrossProd \bfield_1, \nonumber \\
    \pdiff{\ddirds_2}{\dir} & = \qop_2 \left\{ \left(\identityMatrix{3} + \frac{\stepsize}{2} \pdiff{\ddirds_1}{\dir} \right) \columnwiseCrossProd \bfield_2 - \left[ \pdiff{\bfield_2}{\pos_2} \left( \frac{\stepsize}{2}\identityMatrix{3} + \frac{\stepsize^2}{8} \pdiff{\ddirds_1}{\dir} \right) \right] \columnwiseCrossProd \dir_2 \right\}, \nonumber  \\
    \pdiff{\ddirds_3}{\dir} & = \qop_3 \left\{ \left(\identityMatrix{3} + \frac{\stepsize}{2} \pdiff{\ddirds_2}{\dir} \right) \columnwiseCrossProd \bfield_3 - \left[ \pdiff{\bfield_3}{\pos_3} \left( \frac{\stepsize}{2}\identityMatrix{3} + \frac{\stepsize^2}{8} \pdiff{\ddirds_1}{\dir} \right) \right] \columnwiseCrossProd \dir_3 \right\}, \nonumber \\
    \pdiff{\ddirds_4}{\dir} & = \qop_4 \left\{ \left(\identityMatrix{3} + \stepsize \pdiff{\ddirds_3}{\dir} \right) \columnwiseCrossProd \bfield_4 - \left[ \pdiff{\bfield_4}{\pos_4} \left( \stepsize\identityMatrix{3} + \frac{\stepsize^2}{2} \pdiff{\ddirds_3}{\dir} \right) \right] \columnwiseCrossProd \dir_4 \right\}.
\end{align}
\endgroup

For $\pdiff{\ddirds_{\indexRK}}{\qop}$, we need to consider the term with 
$\pdiff{\qop_{\indexRK}}{\qop}$:
\begin{equation}
\pdiff{\ddirds_{\indexRK}}{\qop} = \pdiff{\qop_{\indexRK}}{\qop} (\dir_{\indexRK} \times \bfield_{\indexRK}) + \qop_{\indexRK} \left[ \pdiff{\dir_{\indexRK}}{\qop} \times \bfield_{\indexRK} - \left( \pdiff{\bfield_{\indexRK}}{\pos_{\indexRK}}\pdiff{\pos_{\indexRK}}{\qop} \right) \times \dir_{\indexRK} \right],
\end{equation}
where $\pdiff{\qop_{\indexRK}}{\qop}$ can be expanded using \cref{eq:qop_stages}:
\begin{align}\label{eq:dkndqop}
    \pdiff{\ddirds_1}{\qop} & = \dir_1 \times \bfield_1, \nonumber \\ 
    \pdiff{\ddirds_2}{\qop} & = \left(1 +\frac{\stepsize}{2}\pdiff{\diffquote{\qop_1}}{\qop} \right)(\dir_2 \times \bfield_2) \nonumber \\ & \quad + \qop_2 \left[ \frac{\stepsize}{2}\pdiff{\ddirds_1}{\qop} \times \bfield_2 - \frac{\stepsize^2}{8}\left( \pdiff{\bfield_2}{\pos_2} \pdiff{\ddirds_1}{\qop} \right) \times \dir_2  \right], \nonumber \\
    \pdiff{\ddirds_3}{\qop} & = \left(1 +\frac{\stepsize}{2}\pdiff{\diffquote{\qop_2}}{\qop} \right)(\dir_3 \times \bfield_3) \nonumber \\ & \quad + \qop_3 \left[ \frac{\stepsize}{2}\pdiff{\ddirds_2}{\qop} \times \bfield_3 - \frac{\stepsize^2}{8}\left( \pdiff{\bfield_3}{\pos_3} \pdiff{\ddirds_1}{\qop} \right) \times \dir_3  \right], \nonumber \\
    \pdiff{\ddirds_4}{\qop} & = \left(1 +\stepsize\pdiff{\diffquote{\qop_3}}{\qop} \right)(\dir_4 \times \bfield_4) \nonumber \\ & \quad + \qop_4 \left[ \stepsize\pdiff{\ddirds_3}{\qop} \times \bfield_4 - \frac{\stepsize^2}{2}\left( \pdiff{\bfield_4}{\pos_4} \pdiff{\ddirds_3}{\qop} \right) \times \dir_4  \right].
\end{align}
$\pdiff{\diffquote{\qop_{\indexRK}}}{\qop}$ can be represented as $\pdiff{\diffquote{\qop_{\indexRK}}}{\qop_{\indexRK}}\pdiff{\qop_{\indexRK}}{\qop}$ by the chain rule, thus, can be obtained recursively:
\begingroup
\addtolength{\jot}{\equationSpacingB}
\begin{align}\label{eq:d2qopdsdqop}
 \pdiff{\diffquote{\qop_{1}}}{\qop} & = \pdiff{\diffquote{\qop_{1}}}{\qop_1}, \nonumber \\
\pdiff{\diffquote{\qop_{2}}}{\qop} & = \pdiff{\diffquote{\qop_{2}}}{\qop_2}\left( 1 + \frac{\stepsize}{2}\pdiff{\diffquote{\qop_{1}}}{\qop} \right), \nonumber \\
\pdiff{\diffquote{\qop_{3}}}{\qop} & = \pdiff{\diffquote{\qop_{3}}}{\qop_3}\left( 1 + \frac{\stepsize}{2}\pdiff{\diffquote{\qop_{2}}}{\qop} \right), \nonumber \\
\pdiff{\diffquote{\qop_{4}}}{\qop} & = \pdiff{\diffquote{\qop_{4}}}{\qop_4}\left( 1 + \stepsize\pdiff{\diffquote{\qop_{3}}}{\qop} \right),
\end{align}
\endgroup
where \pdiff{\diffquote{\qop_{\indexRK}}}{\qop_{\indexRK}} is given as follows~\cite{LUND2009_CovarianceTransport}:
\begin{equation}
    \pdiff{\diffquote{\qop_{\indexRK}}}{\qop_{\indexRK}} = \dqopds_{\indexRK} \left[\frac{1}{\qop_{\indexRK}} \left(3 - \frac{\momentum_{\indexRK}^2}{\energy_{\indexRK}^2} \right) + \left(-\pdiff{\energy_{\indexRK}}{\pathlength}\right)^{-1} \pdiff{}{\qop}\left(-\pdiff{\energy_{\indexRK}}{\pathlength}\right) \right].
\end{equation}
In \ref{appendix:mean_eloss}, the analytic solution of $\pdiff{}{\qop}\left(\stoppingPower\right)$ is derived for the Bethe equation.

As for $\pdiff{\qop_{\indexRK}}{\pos}$, $\pdiff{\diffquote{\qop_{\indexRK}}}{\pos}$ and $\pdiff{\diffquote{\qop_{\indexRK}}}{\dir}$ can be neglected for the homogeneous materials, therefore, the derivatives of $\newstate{\qop}$ with respect to $\pos$ and  $\dir$ are zero matrices:
\begin{align}
    \pdiff{\newstate{\qop}}{\pos} & = \zeroMatrix{1}{3}, \\
    \pdiff{\newstate{\qop}}{\dir} & = \zeroMatrix{1}{3}. 
\end{align}
Finally, $\pdiff{\newstate{\qop}}{\qop}$ can be calculated from \cref{eq:d2qopdsdqop}.

In case the gradients of the magnetic field are small enough, the $\pdiff{\bfield_\indexRK}{\pos_\indexRK}$ terms of \cref{eq:dknddir,eq:dkndqop} can be neglected for the fast software implementation.

\section{Solution of $\pdiff{\localFinal}{\dirFinal}$ and $\pdiff{\posInitial}{\azimuInitial}$ in the perigee frame}\label{appendix:perigee_proof}

In this section, we prove that $\pdiff{\localFinal}{\dirFinal}$ is a zero matrix and derive $\pdiff{\posInitial}{\azimuInitial}$ of \cref{eq:drdphi} in a complete manner.  

\subsection{Proof of $\pdiff{\localFinal}{\dirFinal} = \zeroMatrix{2}{3}$}

The subscripts of $\final$ will be omitted because the derivation is valid for any frame intersection. We can start by calculating the derivative of \cref{eq:mu_in_uv} with respect to $\dirFinal$: 
\begin{equation}\label{eq:dmudt}
    \pdiff{\local}{\dir} = 
    \begin{pmatrix}[\matrixSpacingDefault]
    \pdiff{}{\dir}\left((\pos - \frameCenter) \cdot \ubasis \right) \\
    \pdiff{}{\dir}\left((\pos - \frameCenter) \cdot \vbasis \right)
    \end{pmatrix} =
    \begin{pmatrix}[\matrixSpacingDefault]
    \transpose{(\pos - \frameCenter)} \pdiff{}{\dir} \left( \frac{\vbasis \times \dir}{|\vbasis \times \dir|} \right) \\
    \zeroMatrix{1}{3}
    \end{pmatrix}.    
\end{equation}

$\pdiff{}{\dir} \left( \frac{\vbasis \times \dir}{|\vbasis \times \dir|} \right)$ can be expanded by the chain rule and the column-wise cross product, $\columnwiseCrossProd$, defined in \cref{eq:columnwise_cross_product}:
\begin{align}\label{eq:dudt}
    \pdiff{}{\dir} \left( \frac{\vbasis \times \dir}{|\vbasis \times \dir|} \right) & = - \frac{1}{|\vbasis \times \dir|} \identityMatrix{3} \columnwiseCrossProd \vbasis - \frac{\vbasis \times \dir}{|\vbasis \times \dir|^2}\pdiff{|\vbasis\times\dir|}{\dir} \nonumber \\
    & = - \frac{1}{|\vbasis \times \dir|} \left( \identityMatrix{3} \columnwiseCrossProd \vbasis + \ubasis \pdiff{|\vbasis\times\dir|}{\dir} \right).
\end{align}
$\pdiff{|\vbasis \times \dir|}{\dir}$ can be detailed as follows:
\begin{align}\label{eq:dvtdt}
\pdiff{|\vbasis\times\dir|}{\dir} & = \pdiff{\sqrt{(\vbasis \times \dir)\cdot(\vbasis \times \dir)}}{\dir} \nonumber \\
& = \frac{1}{2|\vbasis \times \dir|} \pdiff{[(\vbasis \times \dir)\cdot(\vbasis \times \dir)]}{\dir} \nonumber \\
& = \frac{1}{|\vbasis \times \dir|}  \transpose{(\vbasis \times \dir)} \pdiff{(\vbasis \times \dir)}{\dir} \nonumber \\
& = -\transpose{\ubasis} (\identityMatrix{3} \columnwiseCrossProd \vbasis).
\end{align}

Using \cref{eq:dudt,eq:dvtdt}, we can show that $\pdiff{\ulocal}{\dir}$ of \cref{eq:dmudt} is also the zero matrix:
\begin{align}
\transpose{(\pos - \frameCenter)}\pdiff{}{\dir} \left( \frac{\vbasis \times \dir}{|\vbasis \times \dir|} \right) & =
-\frac{\transpose{(\pos - \frameCenter)} }{|\vbasis \times \dir|}(\identityMatrix{3} - \ubasis \transpose{\ubasis})(\identityMatrix{3} \columnwiseCrossProd \vbasis) \nonumber \\
& = - \frac{\transpose{(\ulocal \ubasis + \vlocal \vbasis)} (\identityMatrix{3} - \ubasis \transpose{\ubasis}) }{|\vbasis \times \dir|} (\identityMatrix{3} \columnwiseCrossProd \vbasis)  \nonumber \\
& = -\frac{\vlocal \transpose{\vbasis}}{|\vbasis \times \dir|}(\identityMatrix{3} \columnwiseCrossProd \vbasis) \nonumber \\
& = \zeroMatrix{1}{3}.
\end{align}

\subsection{Derivation of $\pdiff{\posInitial}{\azimuInitial}$}

We will omit the subscripts of $\initial$ because the derivation is valid for any frame intersection. Here are several equations useful for the derivation:

\begin{align}\label{eq:dvcrosstdphi}
   \pdiff{|\vbasis \times \dir|}{\azimu} & = \pdiff{\sin{\vtangle}}{\azimu} \nonumber \\
   & =  \cos{\vtangle} \pdiff{\vtangle}{\azimu} \nonumber \\
   & =  \vbasis \cdot \dir \pdiff{\cos^{-1}{\vbasis \cdot \dir}}{\azimu} \nonumber \\
    & = - \frac{\vbasis \cdot \dir}{\sqrt{1- (\vbasis \cdot \dir)^2}} \vbasis \cdot \pdiff{\dir}{ \azimu} \nonumber \\ 
   & = - \frac{\vbasis \cdot \dir}{|\vbasis \times \dir|} \vbasis \cdot \pdiff{\dir}{ \azimu},
\end{align}
where $\vtangle$ is an angle between $\vbasis$ and $\dir$ in the range of $[0, \pi]$. 

From the unit vector condition of $(\dir \cdot \dir = 1)$, it is easy to derive the orthogonality between $\dir$ and its derivative with respect to $\azimu$:

\begin{equation}\label{eq:orthogonality_between_t_dtdphi}
     \dir \cdot \pdiff{\dir}{\azimu} = 0.
\end{equation}

The triple cross product of $\vbasis \times \dir \times \vbasis$ satisfies the following relation:

\begin{equation}\label{eq:vtv}
    \vbasis \times \dir \times \vbasis = \dir - \vbasis(\dir \cdot \vbasis).
\end{equation}

Using the above equations, we can derive $\pdiff{\pos}{\azimu}$ by starting from \cref{eq:r_in_uv}. As the derivatives of the last two terms of \cref{eq:r_in_uv} with respect to $\azimu$ is zero, $\pdiff{\pos}{\azimu}$ can be simplified as follows:

\begin{align}\label{eq:drdphi_1}
    \pdiff{\pos}{\azimu} & = \ulocal \pdiff{\ubasis}{\azimu} \nonumber
    \\ & = \ulocal \pdiff{}{\azimu} \frac{\vbasis \times \dir}{|\vbasis \times \dir|} \nonumber \\
    & = \ulocal \left[ - \frac{\vbasis \times \dir}{|\vbasis \times \dir|^2} \pdiff{|\vbasis \times \dir|}{\azimu} + \frac{1}{|\vbasis \times \dir|} \vbasis \times \pdiff{\dir}{\azimu} \right].
\end{align}
Application of \cref{eq:dvcrosstdphi} to the above equation leads to the following: 

\begin{equation}\label{eq:drdphi_2}
 \pdiff{\pos}{\azimu} = \frac{\ulocal}{|\vbasis \times \dir|}\left[ - \ubasis \left( -\frac{\vbasis \cdot \dir}{|\vbasis \times \dir|} \vbasis \cdot \pdiff{\dir}{\azimu} \right)  + \vbasis \times \pdiff{\dir}{\azimu}  \right]. 
\end{equation}
We can add a dummy term using \cref{eq:orthogonality_between_t_dtdphi} and further simplify the expression using \cref{eq:vtv}:

\begin{align}\label{eq:drdphi_3}
 \pdiff{\pos}{\azimu} & = \frac{\ulocal}{|\vbasis \times \dir|}\left[ - \ubasis \left( \frac{ (\dir-(\vbasis \cdot \dir) \vbasis)}{|\vbasis \times \dir|} \cdot \pdiff{\dir}{\azimu} \right)  + \vbasis \times \pdiff{\dir}{\azimu}  \right] \nonumber \\
 & = \frac{\ulocal}{|\vbasis \times \dir|} \left[ - \ubasis \left(  \frac{\vbasis \times \dir}{|\vbasis \times \dir|} \times \vbasis \cdot \pdiff{\dir}{\azimu} \right) + \vbasis \times \pdiff{\dir}{\azimu}  \right] \nonumber \\
& = \frac{\ulocal}{|\vbasis \times \dir|} \left[ - \ubasis \left(  \ubasis \times \vbasis \cdot \pdiff{\dir}{\azimu} \right) + \vbasis \times \pdiff{\dir}{\azimu}  \right].
\end{align}
The derivative can be finalized by permuting the three vectors in the round brackets.
\begin{equation}\label{eq:drdphi_4}
\drdphi.
\end{equation}

\section{Mean energy loss from the Bethe equation and its derivative with respect to $\qop$}\label{appendix:mean_eloss}

In this section, the equations for the mean energy loss from the Bethe equation are enumerated \cite{WORKMAN2022_PDG} and their derivatives with respect to $\qop$ are solved. 

\begin{table}[ht]
\caption{Definitions of variables for the Bethe equation.}
\label{tab:eloss_variables}
\begin{center}
\begin{tabular}{c c l}
 \toprule
 Type & Symbol & \multicolumn{1}{c}{Definition} \\
 \midrule
 \multirowcell{2}{constant} & $\Kcoeff$ & \SI{0.307075}{\MeV \cm^2 \per\mol} \\
          & $\mass_e$ & electron mass of \SI{0.511}{\MeV/c^2} \\
 \midrule
 \multirowcell{4}{absorber \\ material \\ property} & $\density$ & mass density \\
          & $\meanExcitE$ & mean excitation energy \\ 
          & $\atomicNumber$ & atomic number \\
          & $\massNumber$ & atomic molar mass \\
 \midrule
 \multirowcell{4}{incident \\ particle \\ property} 
          & $\chargeNumber$ & charge number \\
          & $\incidentMass$ & mass \\
          & $\beta$ & the ratio of speed and $c$ \\  
          & $\gamma$ & Lorentz factor \\  
 \midrule
          & \multirowcell{2}{$\maxTransE$} & \multirowcell{2}{maximum possible energy transfer \\ to an electron in a single collision} \\
    etc.  &              & \\ 
          & $\delta(\beta\gamma)$ & density effect correction \\
 \bottomrule
\end{tabular}    
\end{center}
\end{table}

\subsection{Mean energy loss from the Bethe equation}

The mean energy loss per path length of charged particles which are relativistic ($0.1 \lesssim \beta\gamma \lesssim 1000$) and heavy is described by the Bethe equation \cite{WORKMAN2022_PDG, BETHE1930_Bethe}:
\begin{equation}\label{eq:bethe}
    \left<\stoppingPower\right> = \Kcoeff \frac{\atomicNumber}{\massNumber}\frac{\density \chargeNumber^2}{\beta^2}\left[ \frac{1}{2} \ln{\frac{2\mass_e c^2 \beta^2 \gamma^2 \maxTransE}{\meanExcitE^2}} - \beta^2 - \frac{\delta(\beta \gamma)}{2} \right],
\end{equation}
where the definitions of variables are listed in \cref{tab:eloss_variables}. The material properties of $\density$, $\meanExcitE$, $\atomicNumber$, and $\massNumber$ can be found in~\cite{GROOM2021_NUCLEARTABLE}. $\maxTransE$ is computed by the following equation:
\begin{equation}
    \maxTransE = \frac{2 \mass_e c^2 \beta^2 \gamma^2}{ 1 + 2\gamma \mass_e /\incidentMass + (\mass_e /\incidentMass)^2}.
\end{equation}
The density effect correction is obtained by \cite{STERNHEIMER1952_DED}:
\begin{equation}
  \delta(\beta\gamma) =
    \begin{cases}
      2(\ln{10})\logbetagamma - \widebar{C}      & \textrm{if } \logbetagamma \geq x_1 \\
      2(\ln{10})\logbetagamma - \widebar{C} + a (x_1 - \logbetagamma)^k & \textrm{if } x_0 \leq \logbetagamma < x_1 \\
      0                                   & \textrm{if } \logbetagamma < x_0  \textrm{ (nonconductors)} \\
      \delta_0 10^{2(\logbetagamma-x_0)}  & \textrm{if } \logbetagamma < x_0  \textrm{ (conductors)}
    \end{cases}   
\end{equation}
where $\logbetagamma$ is the parametrization of $\log_{10}{(\beta\gamma)}$, and the density effect data of ($x_0$, $x_1$, $\widebar{C}$, $a$, $k$, and $\delta_0$) for each material can also be found in~\cite{GROOM2021_NUCLEARTABLE}.

\subsection{Derivation of $\pdiff{}{\qop}\left<\stoppingPower\right>$}

The derivative of $\beta$ with respect to $\qop$ is as follows:
\begin{align}\label{eq:dbetadqop}
    \pdiff{\beta}{\qop} & = \pdiff{\beta}{\momentum}\pdiff{\momentum}{\qop} \nonumber \\
    & = \pdiff{}{\momentum}\left( \frac{\momentum/(\incidentMass c)}{\sqrt{1+\momentum^2 / (\incidentMass c)^2}} \right) \pdiff{(\charge/\qop)}{\qop} \nonumber \\
    & = -\frac{1}{\mass c \gamma (1+\beta^2 \gamma^2)}\frac{\charge}{\qop^2} \nonumber \\
    & = -\frac{\beta}{\qop \gamma^2}.
\end{align}
Using the above equation, it is trivial to obtain the derivatives of $\gamma$ and $\beta\gamma$: 
\begin{equation}\label{eq:dgammadqop}
\pdiff{\gamma}{\qop} = \beta \gamma^3 \pdiff{\beta}{\qop} = - \frac{\beta^2 \gamma}{\qop}.
\end{equation}

\begin{equation}\label{eq:dbetagammadqop}
\pdiff{(\beta\gamma)}{\qop} = \gamma^3 \pdiff{\beta}{\qop} = -\frac{\beta\gamma}{\qop}.
\end{equation}

The derivative of $\maxTransE$ is calculated using \cref{eq:dgammadqop,eq:dbetagammadqop}:
\begin{align}\label{eq:dWmaxdqop}
    \pdiff{\maxTransE}{\qop} & = \frac{4\mass_e c^2 \beta\gamma \pdiff{(\beta\gamma)}{\qop}}{1 + 2\gamma \frac{\mass_e}{\incidentMass} + \left(\frac{\mass_e}{\incidentMass}\right)^2} 
    - \frac{4\mass_e c^2 \beta^2 \gamma^2 \left(\frac{\mass_e}{\incidentMass}\right) \pdiff{\gamma}{\qop}}{\left[1 + 2\gamma \frac{\mass_e}{\incidentMass} + \left(\frac{\mass_e}{\incidentMass}\right)^2 \right]^2} \nonumber\\
    & = -\frac{\maxTransE}{\qop} \left( 2 - \frac{\maxTransE}{\gamma \incidentMass c^2} \right).
\end{align}

The derivative of the Bethe equation with respect to $\qop$ is calculated using \cref{eq:dbetadqop,eq:dbetagammadqop,eq:dWmaxdqop}:
\begin{align}
    \pdiff{}{\qop}\left<\stoppingPower\right> & = \frac{2}{\qop \gamma^2}  \left<\stoppingPower\right>  \nonumber \\
     & \quad + \Kcoeff \frac{\atomicNumber}{\massNumber} \frac{\density \chargeNumber^2}{\beta^2} \left[ - \frac{1}{2\qop} \left( 4 - \frac{\maxTransE}{\gamma \incidentMass c^2} \right) +\frac{2\beta^2}{\qop \gamma^2} -\frac{1}{2}\pdiff{\delta(\beta\gamma)}{\qop} \right],
\end{align}
where $\pdiff{\delta(\beta\gamma)}{\qop}$ is given by:
\begin{equation}
  \pdiff{\delta(\beta\gamma)}{\qop} =
    \begin{cases}
      - \frac{2}{\qop} 
      & \textrm{if } \logbetagamma \geq x_1 \\
      - \frac{2}{\qop} + \frac{ak}{\qop\ln{10}} (x_1 - \logbetagamma)^{k-1} 
      & \textrm{if } x_0 \leq \logbetagamma < x_1  \\
      0                   
      & \textrm{if } \logbetagamma < x_0  \textrm{ (nonconductors)} \\
      -\frac{2}{\qop}\delta_0 10^{2(\logbetagamma-x_0)}   
      & \textrm{if } \logbetagamma < x_0  \textrm{ (conductors)}
    \end{cases}   
\end{equation}

\bibliographystyle{elsarticle-num} 
\bibliography{references}

\begin{thebibliography}{10}
\expandafter\ifx\csname url\endcsname\relax
  \def\url#1{\texttt{#1}}\fi
\expandafter\ifx\csname urlprefix\endcsname\relax\def\urlprefix{URL }\fi
\expandafter\ifx\csname href\endcsname\relax
  \def\href#1#2{#2} \def\path#1{#1}\fi

\bibitem{KALMAN1960_Kalman}
R.~E. Kalman, {A New Approach to Linear Filtering and Prediction Problems}, Journal of Basic Engineering 82~(1) (1960) 35--45.

\bibitem{FRUHWIRTH1987_Kalman}
R.~Frühwirth, {Application of Kalman filtering to track and vertex fitting}, Nucl. Instr. and Meth. A 262~(2) (1987) 444--450.

\bibitem{FRUHWIRTH2021_PatternRecognition}
R.~Frühwirth, A.~Strandlie, {Pattern Recognition, Tracking and Vertex Reconstruction in Particle Detectors}, Springer Cham, 2021.

\bibitem{WITTEK1980_Helix}
W.~Wittek, {Transformation of error matrices for different sets of variables which describe a particle trajectory in a magnetic field}, Tech. rep., CERN, Geneva (1980).

\bibitem{WITTEK1981_Helix}
W.~Wittek, {Error propagation along a helix}, Tech. rep., CERN, Geneva (1981).

\bibitem{STRANDLIE2006_Helix}
A.~Strandlie, W.~Wittek, {Derivation of Jacobians for the propagation of covariance matrices of track parameters in homogeneous magnetic fields}, Nucl. Instr. and Meth. A 566~(2) (2006) 687--698.

\bibitem{BUGGE1981_RK}
L.~Bugge, J.~Myrheim, {Tracking and track fitting}, Nuclear Instruments and Methods 179~(2) (1981) 365--381.

\bibitem{LUND2009_CovarianceTransport}
E.~Lund, L.~Bugge, I.~Gavrilenko, A.~Strandlie, {Transport of covariance matrices in the inhomogeneous magnetic field of the ATLAS experiment by the application of a semi-analytical method}, JINST 4~(04) (2009) P04016.

\bibitem{LUND2009_RKNPropagation}
E.~Lund, L.~Bugge, I.~Gavrilenko, A.~Strandlie, {Track parameter propagation through the application of a new adaptive Runge-Kutta-Nyström method in the ATLAS experiment}, JINST 4~(04) (2009) P04001.

\bibitem{NYSTROM1925_RKN}
E.~Nystr{\"o}m, {{\"U}ber die numerische Integration von Differentialgleichungen}, Acta Societatis scientiarum Fennicae, Druck der Finnischen literaturgesellschaft, 1925.

\bibitem{BILLOIR1992_Perigee}
P.~Billoir, S.~Qian, {Fast vertex fitting with a local parametrization of tracks}, Nucl. Instr. and Meth. A 311~(1) (1992) 139--150.

\bibitem{WORKMAN2022_PDG}
R.~L. Workman, et~al., {Review of Particle Physics}, PTEP 2022 (2022) 083C01.

\bibitem{GESSINGER2023_ODDACAT}
P.~Gessinger-Befurt, A.~Salzburger, J.~Niermann, {The Open Data Detector Tracking System}, Journal of Physics: Conference Series 2438~(1) (2023) 012110.

\bibitem{CORENTIN2022_ODD}
C.~Allaire, P.~Gessinger, J.~Hdrinka, M.~Kiehn, F.~Kimpel, J.~Niermann, A.~Salzburger, S.~Sevova, {OpenDataDetector} (Apr. 2022).
\newblock \href {https://doi.org/10.5281/zenodo.6445359} {\path{doi:10.5281/zenodo.6445359}}.

\bibitem{GROOM2021_NUCLEARTABLE}
D.~E. Groom, et~al., {Atomic and Nuclear Properties of Materials}, \url{https://pdg.lbl.gov/2023/AtomicNuclearProperties/index.html} (2023).

\bibitem{SWATMAN2023_covfie}
S.~N. Swatman, A.-L. Varbanescu, A.~Pimentel, A.~Salzburger, A.~Krasznahorkay, {Systematically Exploring High-Performance Representations of Vector Fields Through Compile-Time Composition}, in: {Proceedings of the 2023 ACM/SPEC International Conference on Performance Engineering}, ICPE '23, Association for Computing Machinery, New York, NY, USA, 2023, p. 55–66.

\bibitem{BETHE1930_Bethe}
H.~Bethe, {Zur Theorie des Durchgangs schneller Korpuskularstrahlen durch Materie}, Annalen der Physik 397~(3) (1930) 325--400.

\bibitem{PRESS2007_Recipe}
W.~H. Press, S.~A. Teukolsky, W.~T. Vetterling, B.~P. Flannery, {Numerical Recipes 3rd Edition: The Art of Scientific Computing}, 3rd Edition, Cambridge University Press, 2007.

\bibitem{SALZBURGER2023_detray}
A.~Salzburger, J.~Niermann, B.~Yeo, A.~Krasznahorkay, {Detray: a compile time polymorphic tracking geometry description}, Journal of Physics: Conference Series 2438~(1) (2023) 012026.

\bibitem{RIDDERS1982_DIFF}
C.~Ridders, {Accurate computation of F$'$(x) and F$'$(x) F$"$(x)}, Advances in Engineering Software (1978) 4~(2) (1982) 75--76.

\bibitem{NEVILLE1934_Extrapolation}
E.~H. Neville, {Iterative interpolation}, St. Joseph's IS Press, 1934.

\bibitem{ATLAS_TDR}
{Technical Design Report for the ATLAS Inner Tracker Pixel Detector}, Tech. rep., CERN, Geneva (2017).

\bibitem{MURDOCH2008_PVALUE}
Y.-L.~T. Duncan J~Murdoch, J.~Adcock, {P-Values are Random Variables}, The American Statistician 62~(3) (2008) 242--245.
\newblock \href {https://doi.org/10.1198/000313008X332421} {\path{doi:10.1198/000313008X332421}}.

\bibitem{Landau1944_EnergyLoss}
L.~D. Landau, {On the energy loss of fast particles by ionization}, J. Phys. 8~(4) (1944) 201--205.

\bibitem{Vavilov1957_EnergyLoss}
P.~V. Vavilov, {Ionization losses of high-energy heavy particles}, Sov. Phys. JETP 5 (1957) 749--751.

\bibitem{BICHSEL1988_MostProbableLoss}
H.~Bichsel, Straggling in thin silicon detectors, Rev. Mod. Phys. 60 (1988) 663--699.

\bibitem{HIGHLAND1975_MS}
V.~L. Highland, {Some practical remarks on multiple scattering}, Nuclear Instruments and Methods 129~(2) (1975) 497--499.

\bibitem{LYNCH1991_MS}
G.~R. Lynch, O.~I. Dahl, {Approximations to multiple Coulomb scattering}, Nuclear Instruments and Methods in Physics Research Section B: Beam Interactions with Materials and Atoms 58~(1) (1991) 6--10.

\bibitem{FRUHWIRTH2000_DataAnalysis}
R.~Frühwirth, M.~Regler, R.~K. Bock, H.~Grote, D.~Notz, {Data analysis techniques for high-energy physics; 2nd ed.}, Cambridge monographs on particle physics, nuclear physics, and cosmology, Cambridge Univ. Press, Cambridge, 2000.

\bibitem{ROSSI1952_Book}
B.~B. Rossi, {High-energy particles}, Prentice-Hall physics series, Prentice-Hall, New York, NY, 1952.

\bibitem{FRUHWIRTH2003_GSF}
R.~Frühwirth, {A Gaussian-mixture approximation of the Bethe–Heitler model of electron energy loss by bremsstrahlung}, Computer Physics Communications 154~(2) (2003) 131--142.

\bibitem{HESSEY2013_Annulus}
N.~Hessey, {Building a Stereo-angle into strip-sensors for the ATLAS-Upgrade Inner-Tracker Endcaps}, Tech. rep., CERN, Geneva (2013).

\bibitem{ABRAMOWITZ1972_Handbook}
M.~Abramowitz, I.~A. Stegun (Eds.), {Handbook of Mathematical Functions with Formulas, Graphs, and Mathematical Tables}, tenth printing Edition, U.S. Government Printing Office, Washington, DC, USA, 1972.

\bibitem{STERNHEIMER1952_DED}
R.~M. Sternheimer, {The Density Effect for the Ionization Loss in Various Materials}, Phys. Rev. 88 (1952) 851--859.

\end{thebibliography}






\end{document}